\documentclass[letterpaper,10 pt,conference]{ieeeconf}
\IEEEoverridecommandlockouts

\usepackage[dvipsnames]{xcolor}
\usepackage{cite}
\usepackage{amsmath,amssymb,amsfonts}
\usepackage{graphicx}
\usepackage{textcomp}
\usepackage{comment}
\usepackage{subcaption}
\def\BibTeX{{\rm B\kern-.05em{\sc i\kern-.025em b}\kern-.08em
    T\kern-.1667em\lower.7ex\hbox{E}\kern-.125emX}}
\usepackage{float}
\let\labelindent\relax
\usepackage{enumitem}
\usepackage[font=small]{caption}

\usepackage{amsthm}
\usepackage{amssymb}
\usepackage[hidelinks]{hyperref}
\newtheorem{definition}{Definition}

\newtheorem{lemma}{Lemma}
\newtheorem{prop}{Proposition}
\newtheorem{theorem}{Theorem}
\usepackage{color,soul}
\newtheorem{remark}{Remark}
\newtheorem{result}{Result}
\usepackage{algpseudocode}
\usepackage{mathtools} \mathtoolsset{showonlyrefs}

\usepackage{url}
\usepackage{enumitem}
\setlist[enumerate]{leftmargin=*}
\setlist[itemize]{leftmargin=*}

\usepackage{tabularx}
\usepackage{booktabs} 

\def\baselinestretch{0.985}

\title{\huge {\color{black}The PenduMAV: A Six-Input Omnidirectional MAV without Internal Forces - Design, Dynamics, and SE(3) Control}}

\author{Ahmed Ali$^{*}$, Quentin Sablé$^{*}$, Chiara Gabellieri$^{*}$, Antonio Franchi$^{*,\dagger}$
\thanks{$^*$ Robotics and Mechatronics Department, Electrical Engineering,  Mathematics, and Computer Science (EEMCS) Faculty, University of Twente, 7500 AE Enschede, The Netherlands. {\footnotesize ahmed.ali@utwente.nl, q.l.g.sable@utwente.nl, c.gabellieri@utwente.nl, schol@r-franchi.eu}}
\thanks{$^\dagger$Department of Computer, Control and Management Engineering, Sapienza University of Rome, 00185 Rome, Italy. {\footnotesize schol@r-franchi.eu}}\thanks{This work was partially funded by the Horizon Europe research agreement no. 101120732 (AUTOASSESS).}}
\IEEEoverridecommandlockouts 
\begin{document}

\maketitle


\begin{abstract}
 We {\color{black} introduce the PenduMAV, an exactly actuated (6-input) omnidirectional multirotor that structurally eliminates internal forces at equilibria. The vehicle features one actively tilting propeller and three propellers mounted on passive pendulum links via universal joints. This architecture achieves full 6D wrench generation while avoiding the structural and energetic costs of  input redundancy and internal forces. After deriving the full multibody dynamics, we demonstrate that a forced equilibrium exists for every main platform pose. To asymptotically stabilize the closed-loop system, we design a coordinate-invariant nonlinear controller based on dynamic feedback linearization and backstepping, utilizing the left-trivialized error on $SE(3)$. System stability is formally guaranteed through Lyapunov analysis of the zero dynamics. Finally, Gazebo simulations validate the approach, showcasing fully decoupled attitude and translational tracking under different uncertainties and actuator noise.}
\end{abstract}

\section{introduction}
\label{sec:intro}

 Interest in Multirotor Aerial Vehicles (MAVs) is growing, driven by its potential to perform tasks in environments inaccessible to other  mobile vehicles~\cite{sabour2023applications, antonioReview}. 
Applications such as contact-based inspection of sloped surfaces~\cite{slide,FullyInspection} have emerged, in which the capability of executing decoupled translational and rotational motions is essential, necessitating the demand for new MAV designs since quadrotor {\color{black}{dynamics inherently prohibit}} this kind of motion.

\subsubsection*{Research Problems and Related Work}

One can easily be convinced that it is desirable for a MAV to have all the angular velocities of the propellers aligned with each other and with the gravity force at any static hovering equilibrium: this avoids internal propeller-generated forces canceling each other at the cost of a higher energy consumption; moreover, it is desirable to keep the number of actuators to the minimum (equal to the task dimension), to reduce the {\color{black}weight} (which, in turn, affects the vehicle's endurance), the cost, the maintenance requirements, etc. Furthermore, for tasks requiring high dexterity, it is desirable for the MAV to be able to statically hover at any pose (in SE(3) in general). Last but not least, the design of a MAV must also allow the controlled stabilization of its hovering equilibria.
We label these four desirable properties as the Internal-force Property, the Input-dimension Property, the Equilibria Property, and the Stabilizability Property, respectively. 

When restricting the problem to the vertical 2D space, we demonstrated in~\cite{ali2024controltheoreticstudyomnidirectional} that a MAV design with all the above four properties exists. 
 However, extending such a result to the 3D space is not trivial. Just to mention two obstacles to such an extension:  SE(3) is a fundamentally different manifold than SE(2) and the dynamic effects related to single revolute joints in 2D cannot be extended to the case of multiple joints in 3D, see model in Sec.~\ref{sec:eqmotion}. Moreover, in the 3D space, propeller drag torques arise, see discussion in Sec.~\ref{SecSpring}.

\subsubsection*{State of the Art}
To the best of the authors' knowledge, no MAV designs proposed in the literature have all the four desirable properties described above.

Standard underactuated MAVs tick the Internal-force Property as they have parallel propellers; however, the only hovering equilibria are those where the $z$ axis is aligned with gravity. Thus, they do not tick the Equilibria property.

{\color{black} MAVs} equipped with the ability to hover while maintaining a larger set of orientations, known as fully-actuated MAVs, are surveyed in~\cite{FullyReview}. These MAVs cannot be stabilized at any pose but at a subset of poses in SE(3) determined by the thrust limits, the design kinematics, and joint limits. 
A subclass of this type of MAVs, called omnidirectional MAV~\cite{omni}, can hover at \textit{any} pose: namely, they have the Equilibria property mentioned above. Fully-actuated aerial vehicles can be divided in two classes.

The first class is realized by tilted propellers with non-parallel spinning axes. As a consequence, none of these designs has the Internal-force Property. Examples are the fully actuated hexarotor~\cite{TiltHex} and the omnidirectional ones in~\cite{Lynchpin6inputs,ODAR6inputs}. Other omnidirectional fixedly-tilted-propeller designs do not comply with the Input-dimension Property, either; for example, the hepta-rotor of~\cite{hamandi2021understanding} or the octarotors proposed in~\cite{8reversible, ODAR8inputs}.

The second design class is realized by the addition of servomotors that actively change the relative orientation between the propellers and the frame of the MAV.
In~\cite{ryll2014novel,QuadTilt8inputs}, the proposed design separately tilts 4 propellers via one servomotor each, for a total of 8 actuators.
Similarly, in~\cite{Voliro12inputs}, the authors used 6 servomotors in addition to the 6 brushless motors that rotate the 6 propellers; {\color{black}{in \cite{Nguyen2023Adaptive}, 3 servos actively tilt the tricopter's propellers while producing internal force at hovering and without achieving full omnidirectional motion}}.  In~\cite{Dragon}, a MAV with multiple links connected in series by active revolute joints is proposed, with each link having two propellers rotated by one motor. In an attempt to limit the number of actuators, it has been proposed to use a single servo-motor to synchronously tilt all the 6 ~\cite{FastHex} or 8~\cite{aboudorra2024modelling} propellers. {\color{black}{In \cite{Iriarte2021PassiveRotorTilting}, several quadrotor modules are passively tilting  to achieve full actuation}}. All the aforementioned designs do not comply with the Input-dimension Property, having a number of actuators greater than~6. Furthermore they do not comply with the Internal-force Property at any orientation but only some at some specific ones. {\color{black}{While the VTOL in \cite{Ito2024PassiveJointVTOL} has 6 propellers, some of which are passively-tiltable, it is not omnidirectional and generates internal forces at some orientations.}}

\begin{table}[htbp]
\caption{Nomenclature}
\label{tab:nomenclature}
\centering
\footnotesize 
{\color{black}
\begin{tabularx}{\columnwidth}{@{}l X@{}}
\toprule
\textbf{Symbol} & \textbf{Description} \\
\midrule
$SE(3)$ & Special Euclidean group of rigid-body transformations. \\
${se}(3)$ & Lie algebra of $SE(3)$. \\
$\mathbb{T}^n$ & $n$-torus manifold (a compact Lie group). \\
$C^{0}_{j} \in SE(3)$ & Configuration of $B_j$'s frame $\mathcal{F}_{p_j}$ in the world frame $\mathcal{F}_w$. \\
$V^{0}_{1} \in \mathbb{R}^6$ & Right-invariant base twist expressed in, and relative to, $\mathcal{F}_w$. \\
$[*]$ & $\mathbb{R}^6\to se(3)$ An isomorphism with an inverse $\hat{*}$. \\
$(*)_\times$ & The skew-symmetric matrix of the vector $*\in\mathbb{R}^3$. \\
$J_{1j}$ & Universal joint connecting the base to link $B_j$. \\
$S_j \in \mathbb{R}^{6\times 2}$ & $J_{1j}$'s spatial instantaneous screw axis, $S_j = (\,S_{j1}\; S_{j2}\,)$. \\
$S_p$ & Block-diagonal matrix of screws $S_j$, $\operatorname{blkdiag}(S_j)$. \\
$\operatorname{Ad}_{C}$ & $SE(3)\times se(3) \rightarrow se(3)$ Adjoint representation of $SE(3)$. \\
$\operatorname{ad}_{V}$ & $se(3)\times se(3) \rightarrow se(3)$ Adjoint representation of $se(3)$. \\
$R_k(*)$ & Pure rotation about axis $k$ by a angle $*$. \\
$e_i$ & Zero vector with a one at index $i$. \\
$M$ & Generalized mass (inertia) matrix. \\
$M_{bb}, M_{bp}, M_{pp}$ & Base, coupling, and joint sub-matrices of $M$. \\
$h$ & Centrifugal and Coriolis terms. \\
$g_b, g_p$ & Gravitational terms of base and joints, $g=(\,g_b\,\,g_p\,)^T$. \\
$W^{0}_{\mathrm{app,b}}$ & Applied wrench on base's CoM expressed in $\mathcal{F}_w$. \\
$W^{0}_{\mathrm{app,p}}$ & Applied wrench on links $B_2, B_3, B_4$'s CoM in $\mathcal{F}_w$ projected onto respective joints via $S_p$. \\
$\tau_{dr_{i}}$ & Drag torque about propeller $i$'s axis. \\
$\tau_{f_{ji}}$ & Viscous friction at joint $q_{ji}$'s axis. $\bar{\tau}_{f} = \big( 0_{6\times1} \;\; \operatorname{col}(\tau_{f_{ji}}) \big)^T$. \\
$\tau_{s_j}$ & Torsion spring torque applied at joint $q_{j1}$'s axis. \\
$\tau_{dj}$ & Propeller $j$'s drag torque factorized on $q_{j1}$'s axis at equilibrium. \\
$m_b, m_j$ & Mass of base and link $B_j$. \\
$c$ & Length from joints axes intersection point to $B_j$'s CoM. \\
$k_i$ & Drag-to-thrust ratio of propeller $i$. \\
$\mathrm{PJD}_1, \mathrm{PJD}_2$ & Passive-joint designs of type 1 and type 2. \\
$P_{se(3)}(A)$ & Projects a matrix $A\in \mathbb{R}^{4\times4}$ onto $se(3)$. \\
$*_e$ & Variable $*$ evaluated at equilibrium. \\
$*^{[i]}$ & Value of variable $*$ pertaining to joint design PJD$_i$. \\
\bottomrule
\end{tabularx}}
\end{table}

\subsubsection*{Contributions}
The main contribution of this work is to propose the PenduMAV concept, the first MAV design, to the best of the authors' knowledge, accomplishing the following:
\begin{enumerate}
    \item Input-dimension property. The number of actuators in the proposed MAV design is 6, equal to the number of DoFs of its base's pose. There are 4 brushless motors, each making one propeller spin; 3 of the propellers are connected to a passive 2-DoF joint; one of the propeller's attitude is actively controlled by 2 servomotors.
        
        \item Equilibria property.  In Proposition~\ref{Prop1}, we show that static hovering is feasible at \textit{any} pose in $SE(3)$ of the MAV's main body; in other words, a forced equilibrium exists for any base's pose.
        \item Internal-force property. Proposition~\ref{Prop1} also shows that no internal forces are produced at the equilibrium for any base pose in $SE(3)$.
        \item Stabilizability property. We prove in Theorem~\ref{Theorm1} that it is possible to create a state feedback law, given in Proposition~\ref{Prop2}, that asymptotically stabilizes the closed-loop system at any base pose in $SE(3)$, as long as configuration and thrust trajectories of the system remain in a certain neighborhood of that base pose. We observe that the dynamics of the passive propeller joints coincide with the zero dynamics, and we prove its asymptotical stability using the Lyapunov theory. 
    \end{enumerate}
Eventually, simulation results in the presence of non-idealities, such as noise and parameter uncertainties, support the generalizability of the results beyond the standard model used in the theoretical derivations.
{\color{black}{Complete formal proofs and explicit calculation formulas are collected in the appendices of~\cite{Ali2025}}}.


\section{Proposed Design and Dynamical Model}
\label{sec:Model}

In this section, we describe the proposed design and present its dynamical model in state-space form.

\subsection{Motivations Behind the Design Idea}
The simplest MAV design that one could think of that respects the Input-dimension Property and the Internal-force Property (and potentially the Equilibria Property) is the one with 6 propellers, all free to passively tilt like 3D pendulums. However, that design in the 2D case in~\cite{ali2024controltheoreticstudyomnidirectional} turned out to be non-feedback linearizable, as the allocation matrix was structurally rank-deficient at the static equilibria. Based on such a preliminary result, we discarded the all-passive-propeller design. 

An alternative could be to reorient actively \textit{all} four propellers synchronously by means of the two servomotors and a transmission mechanism. However, the design of such a mechanism would bring mechanical complexity and additional payload, cost, and maintenance effort; furthermore, it would not scale well with the size of the vehicle. Instead, the proposed design where the servomotors exert their actuation locally at one individual propeller has the advantage of being modular and simple from a mechanical point of view.

\subsection{Mechanical Design Idea}
\begin{figure}
    \centering
    \includegraphics[width=1\columnwidth]{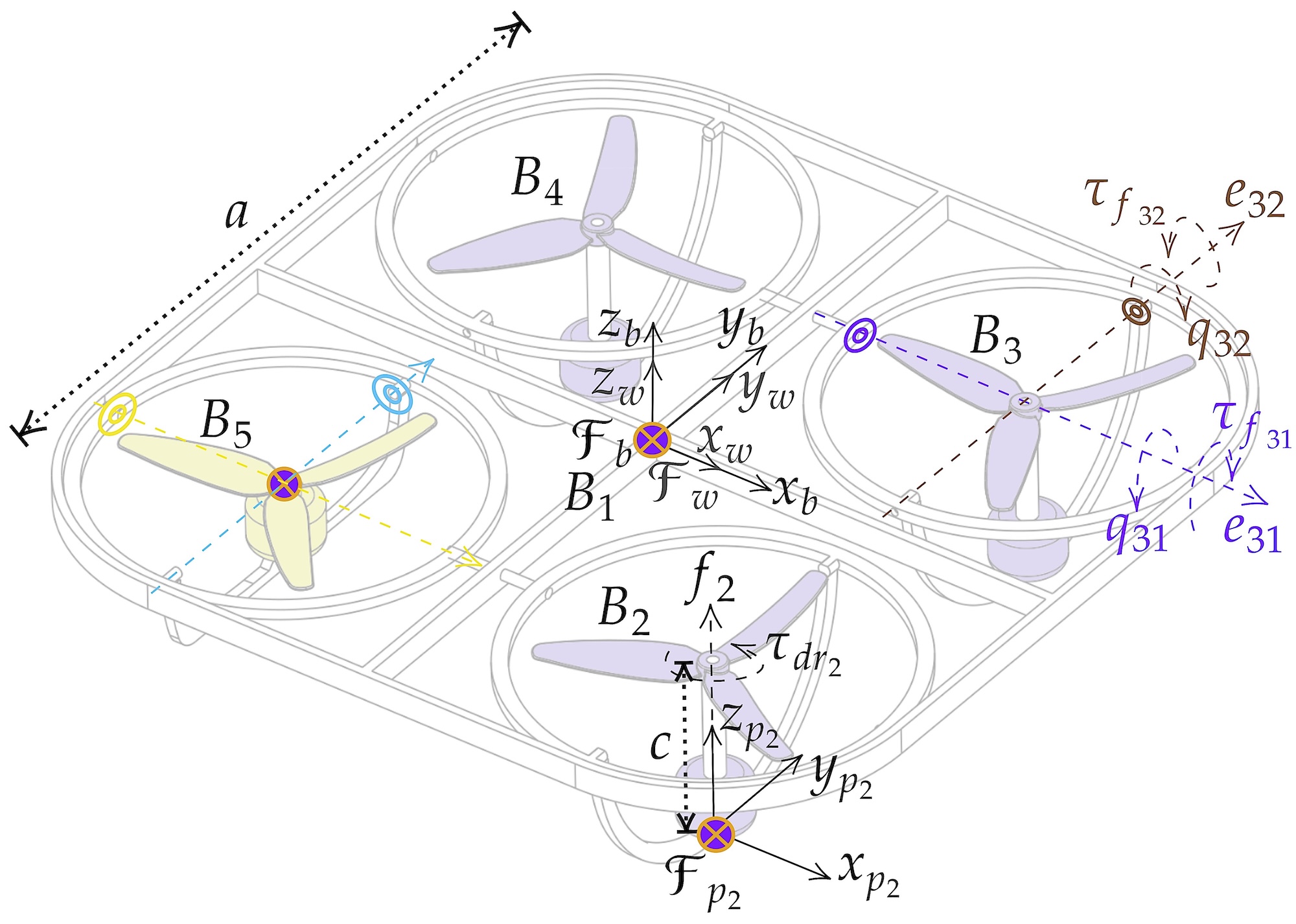}
    \caption{Conceptual representation of the PenduMAV idea. The origins of the frames $\mathcal{F}_b$ and $\mathcal{F}_{p_2}$  are attached to CoMs of the base $B_1$ and the propeller link $B_2$, respectively. The shown configuration is when $\mathcal{F}_b$ coincides with $\mathcal{F}_w$ and all the frames of the links  have the same orientation as $\mathcal{F}_w$. The joint $q_{j1}$ and $q_{j2}$ positions are zero when $\mathcal{F}_{p_j}$ has the same orientation as $\mathcal{F}_b$, $j=\{2,\ldots,5\}$. The link $B_5$ is attached to $B_1$ through an active joint, unlike the the rest of the joints which are passive.}
    \label{main_vehicle}
\end{figure}
\begin{figure}
\centering
    \includegraphics[width=0.5\columnwidth]{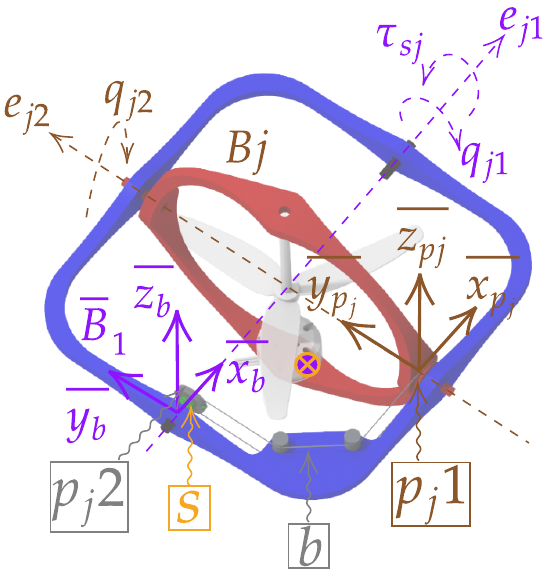}
    \caption{The connection between $B_j$ ($j=2,3,4$) and the base $B_1$ in PJD$_2$ consists of universal passive joint with a Pulley-Belt-Spring module. $\bar{B}_j$ is the intermediate link connecting $B_1$ to $B_j$. $b$, $p$ and $s$ denote the belt, pulley and a torsion spring, respectively. $\tau_{sj}$ is the torsion spring torque around the rotation axis of $q_{j1}$, i.e. $e_{j1}$.
    }   
    \label{fig:universal_spring}
\end{figure}
The PenduMAV concept, shown in Fig.~\ref{main_vehicle}, consists of a main rigid body, referred to as the \emph{base} and denoted by $B_1$, and four smaller bodies, referred to as \emph{links} and denoted by $B_2,B_3,B_4$, and $B_5$. Each link carries a rigidly-attached motor driving a propeller and is connected to the base by two revolute joints. The propeller, the motor, the link, and the two revolute joints together form an \emph{actuation unit}. For each unit, the joints' mutually-orthogonal rotation axes intersect at a point, called the \emph{center of rotation} of the link, coinciding with the propeller center. The actuation units are mounted on $B_1$ such that all centers of rotation lie on the same plane, resulting in a \emph{coplanar design}. The centers of mass (CoMs) of $B_2,B_3$, and $B_4$ are located below their respective centers of rotation, achieved by placing the motors sufficiently low. This makes these links behave as spatial pendula, with the joints acting as pivots. {\color{black}{In contrast, $B_5$'s CoM coincides with its center of rotation}}. $B_1$ is connected to $B_5$ by an actuated universal joint with configuration $q_a=\operatorname{col}(q_{5i})\in\mathbb{T}^2$, and to $B_2,B_3$, and $B_4$ by passive universal joints with configurations $q_p=\operatorname{col}(q_{ji})\in\mathbb{T}^6$, $j=2,\ldots,5$, $i=1,2$.

For modeling simplicity, the actuated joint is assumed to be driven by fast velocity-controlled servo DC motors, neglecting motor dynamics. Moreover, $B_5$'s mass is assumed to be concentrated at its CoM, zeroing its angular inertia moment. Under these assumptions, the projected dynamics of ${B_5}$ onto its $q_a$'s screw axis vanish, rendering $\dot q_a$ governed solely by the motor input. {\color{black}{Although the controller is developed under these assumptions, the realistic simulations in Sec. \ref{sec:Sim} shows that the closed-loop performance remains almost unaffected when these assumptions on $B_5$ are violated.}}

Two Passive Joint Designs (PJD) are considered. In PJD$_1$, the passive joints are implemented as described above. In PJD$_2$, an additional Pulley-Belt-Spring module (Fig.~\ref{fig:universal_spring}) is integrated into each passive joint. This module transfers the rotation about $q_{j2}$'s axis to a rotation about $q_{j1}$'s axis through a belt-pulley transmission. A torsion spring, mounted along $q_{j1}$'s axis, applies a moment proportional to $q_{j2}$, thus the mechanism is unidirectional: rotations about $q_{j2}$ generate a restoring moment about $q_{j1}$, while the reverse is not true. {\color{black}{The rationale behind PJD$_2$ is motivated by a tradeoff between mechanical simplicity and exact internal-force elimination. In PJD$_1$, the drag torque produces a residual torque about the first passive universal-joint axes, slightly shifting the equilibrium away from the ideal zero-internal-force (vertical) configuration. Bringing the passive links back to this ideal configuration at arbitrary base orientation requires compensating this drag-induced residual torque, achieved by the pulley-belt-spring mechanism in PJD$_2$. Therefore, PJD$_2$ should be viewed as an optional refinement that achieves exact internal-force elimination at the cost of additional mechanical complexity.}} Unless otherwise stated, the analysis herein applies to both designs.
\subsection{Equations of Motion}
\label{sec:eqmotion}
We denote with $\mathcal{Q}=(C_1^0,{\color{black}{q_{p}}},{\color{black}{q_a}}) \in SE(3) \times \mathbb{T}^8$ the system's configuration where {\color{black}{\(SE(3)\) denotes the special Euclidean group, i.e., the Lie group of rigid-body poses composed of a rotation \(R \in SO(3)\) and a position \(p \in \mathbb{R}^3\)}}. The velocity of ${\color{black}{(C_1^0,q_{p})}}$ is ${\color{black}{(V_1^0,\dot{q}_{p})}} \in \mathbb{R}^{12}$ where $[V_1^0]=\dot{C}^0_1(C^0_1)^{-1}\in se(3)$. Denote the full state vector $x(t)$ as $x=({\color{black}{C_1^0,q_{p}}},{\color{black}{V_1^0,\dot{q}_{p}}},{\color{black}{q_a}})^T \in \mathbb{X}\equiv SE(3) \times \mathbb{T}^8 \times \mathbb{R}^{12}$. Dropping dependencies for clarity, the system dynamics is thus described by these
state space equations
\begin{equation}
\Sigma_1:\;
\resizebox{0.91\columnwidth}{!}{%
$\left\{
\begin{aligned}
\dot{x} =& {\color{black}{F_1(x)+G_1(x)u}}\\=&
\begin{pmatrix}
[{\color{black}{V_1^0}}]\,{\color{black}{C_1^0}} \\
{\color{black}{\dot{q}_{p}}} \\
M^{-1}(-h - g + \bar{\tau}_{f}) \\
0_{2\times1}
\end{pmatrix}
+
\begin{pmatrix}
0_{12\times6} \\
M^{-1}
\begin{pmatrix}
D_b & 0_{6\times2} \\
{\color{black}{D_p}} & 0_{6\times3}
\end{pmatrix}
\\
0_{2\times4} \;\; I_{2\times2}
\end{pmatrix} u
\end{aligned}
\right.$%
}
\label{statespace}
\end{equation}
where explicit formulas of $M,h$ and $g$ are given in Sec.~\ref{app:ModelDefinitions}~\cite{Ali2025}, and $D_b$, ${\color{black}{D_p}}$ and the input vector $u$ are defined as
\begin{equation}
\resizebox{0.91\columnwidth}{!}{$
    \begin{split}
        D_b=&\left(\begin{array}{cccc}
             \operatorname{Ad}^T_{(C^0_{f_2})^{-1}}\bar{f}_2  & \operatorname{Ad}^T_{(C^0_{f_3})^{-1}}\bar{f}_3 & \operatorname{Ad}^T_{(C^0_{f_4})^{-1}}\bar{f}_4 &
             \operatorname{Ad}^T_{(C^0_{f_5})^{-1}}\bar{f}_5     
        \end{array}\right)_{6\times 4}\\
        {\color{black}{D_p}}=&\left(\begin{array}{ccc}
             S_2^T\operatorname{Ad}^T_{(C^0_{f_2})^{-1}}\bar{f}_2 & 0 & 0 \\ 0 & S_3^T\operatorname{Ad}^T_{(C^0_{f_3})^{-1}}\bar{f}_3& 0 \\ 0&0&S_4^T\operatorname{Ad}^T_{(C^0_{f_4})^{-1}}\bar{f}_4   
        \end{array}\right)_{6\times 3}\\
        u=&\left(\begin{array}{cccccc}
          f_2 & f_3 & f_4 & f_5 & u_{v_1} &u_{v_2}
     \end{array}\right)_{6\times 1}^T \in \mathbb{R}^6
    \end{split}$}
    \label{Db,Dt}
\end{equation}
where $f=col(f_i)$ and $f_i$ is the thrust magnitude of the propeller attached to $B_i$, with $i=2,\ldots,5$; $\Bar{f}_i$ is a constant vector given as
$\bar{f}_i := (\,0\;\;0\;\;k_i\;\;0\;\;0\;\;1\,)_{6\times 1}^T$. $u_v$ is tilting speed. $C^0_{f_i} \in SE(3)$ is the homogeneous transformation of a frame $\mathcal{F}_{f_i}$ attached at the propeller of link $B_i$ relative to the inertial frame $\mathcal{F}_w$.
 
 \begin{result}[Input dimension property]
  The proposed design satisfies the input dimension property. In fact the dimension of the manifold of the configuration of the base frame $C_1^0$ is six and the dimension of the input is $\operatorname{dim}(u)=6$ (four propeller thrusts and two servomotor speeds).
 \end{result}

\section{Study of Attainable Equilibria and\\ Internal Forces at each Equilibrium} 
\label{sec:Equil_Sing}

In this section we obtain the set of forced equilibria  of the dynamics~\eqref{statespace} and investigate their properties in terms of the corresponding four thrust forces produced at these configurations. We begin by giving some preliminary definitions. Afterwards, the main result, {\color{black}{which establishes the Equilibria property and the Internal-force property}}, is summarized in Proposition~\ref{Prop1} whose proof is found in Sec.~\ref{proofProp1} of~\cite{Ali2025}.

Imposing $\dot{x}{(t)}=0$ in~\eqref{statespace} implies $({\color{black}{V_1^0,\dot{q}_{p}}},{\color{black}{\dot{q}_a}})=0$, which defines an equilibrium condition at which the state and the input become, respectively, $x_e=(C^0_{1e},{\color{black}{q_{pe}}},0,0,{\color{black}{q_{ae}}})^T$ and $u_e=(f_{e}\,\,u_{ve})^T$, with $f_e=(f_{2e} \,\, f_{3e} \,\, f_{4e} \,\, f_{5e})^T$ and $u_{ve}=(u_{v_{1e}}\,\, u_{v_{2e}})^T$. Under this condition, both $h({\color{black}{C_1^0,q_{p}}},{\color{black}{V_1^0,\dot{q}_{p}}})=0$ and $\Bar{\tau}_{f}({\color{black}{\dot{q}_{p}}})=0$ vanish, as they depend quadratically and linearly, respectively, on the configuration velocity.

For PJD$_1$, \eqref{statespace} reduces to this static balance vector equation whose admissible solutions are the equilibrium state and input pairs $(x_{e},u_e)$ such that
{\color{black}{\begin{align}
g_b(C^0_{1e},{\color{black}{q_{pe}}})&=(\,\, {\color{black}{D_b}}\,\,0_{6\times 2}\,\,) u_e, \\
    {\color{black}{g_p}}(C^0_{1e},{\color{black}{q_{pe}}})&= (\,\, {\color{black}{D_p}}\,\,0_{6\times 3}\,\,) u_e = (\,\, \tau_{d2}\,\,0\,\,\tau_{d3}\,\,0\,\, \tau_{d4}\,\,0\,\,)_{6\times 1}^T \nonumber\\
    0_2&=u_{ve} \label{static1} 
\end{align}}}   
where $\tau_{dj}=k_{j} f_{je} \sin(q_{j2e})$, with $j=2,3,4$. $C^0_{1e}$ and ${\color{black}{q_{pe}}}=\operatorname{col}(\,q_{je})$, with $q_{je}=\operatorname{col}(\,q_{jie})$ $\forall j=2,3,4, \forall i=1,2$.
\subsection{Effect of the Aerodynamic Drag Torque at the Equilibrium}
\label{SecSpring}
   The presence of the torque $\tau_{dj}$ in~\eqref{static1} makes the value of the joints ${\color{black}{q_{pe}}}$ at the equilibrium is not solely determined by the gravitational torque at equilibrium ${\color{black}{g_{pe}}}$, as it is the case for the 2D MAV presented in~~\cite{ali2024controltheoreticstudyomnidirectional}, but also by a component of the equilibrium thrust $f_{e}$ at equilibrium. 
   
   The residual torque $\tau_{dj}$ can be eliminated at any equilibrium by the spring introduced in PJD$_2$ whose {\color{black}{proposed implementation in Fig. \ref{fig:universal_spring} enables a torsion spring to counteract residual torque and passively return the link to its vertical configuration, zeroing the internal force as stated in proposition \ref{Prop1}}}. {\color{black}{In fact, we can tune each variable-stiffness torsion spring such that its stiffness $k_s=\frac{\partial \tau_{sj}}{\partial q_{j2}}$ varies linearly with $\cos(q_{j2})$, with a constant slope $\bar{k}=k_{j} f_{je}$, i.e. $k_s=\bar{k}\cos(q_{j2})$, which is feasible since $f_{je}$ is found to be constant for any equilibrium of PJD$_2$ as long as the vehicle weight remains constant during flight as shown in Sec.~\ref{proofProp1} of~\cite{Ali2025}. Hence, the spring needs not to change or be re-tuned between flights}}. As a result, $\tau_{dj}$ is canceled at any PJD$_2$ equilibrium, i.e. when $\tau_{dj} = -\tau_{sje}$, transforming~\eqref{static1} to
\begin{equation}
    {\color{black}{g_p}}(C^0_{1e},{\color{black}{q_{pe}}})=0_6,\,\,\, u_{ve}=0_2. \label{Case2static1}
\end{equation}
\subsection{Minimal Thrust at Equilibrium vs Internal Forces}

Before presenting the result on the Equilibria and Internal-force properties of the proposed MAV~\eqref{statespace} we provide two definition for the thrust magnitudes at the equilibrium, in the case in which they are minimal and in the case in which there is internal force.

\begin{definition}
\label{Def1}
 A vector of thrust magnitudes $f\in\mathbb{R}_{\geq 0}^l$, where $\mathbb{R}_{\geq 0}^l := \left\{ s=col(s_i) \in \mathbb{R}^l \mid s_i \geq 0, \ \forall i \in \{1, \dots, l\} \right\}$, is said to be \emph{minimal at the equilibrium} if it belongs to $\mathcal{F}_{\rm min}^{\rm eq}$, defined as: 
\begin{equation}
   \mathcal{F}_{\rm min}^{\rm eq}:=
    \bigg\{ f=col(f_i) \in \mathbb{R}_{\geq 0}^l \bigg|\, \sum^{l}_{i=1}{f}_i= m_{\rm tot}\,g_r \bigg\}\label{minDefinition_Fmin}
\end{equation}
where $m_{\rm tot}$ is the robot's mass and $g_r=9.81 m/s^2$. Instead, $f$ is said to have \emph{internal forces at the equilibrium} if it belongs  to  $\mathcal{F}_{\rm int.for.}^{\rm eq}$, defined as:
\begin{equation}
\mathcal{F}_{\rm int.for.}^{\rm eq}:=
    \bigg\{ f=col(f_i) \in \mathbb{R}_{\geq 0}^l \bigg|\, \sum^{l}_{i=1}{f}_i> m_{\rm tot}\,g_r \bigg\}\label{internalFORCE}
\end{equation}
A MAV design whose thrust magnitudes are minimal at all its equilibria is said to have no internal forces at its equilibria.
\end{definition}
In other words, a minimal thrust magnitudes vector $f_e\in\mathcal{F}_{\rm min}^{\rm eq}$ has the least amount of total thrust magnitudes necessary to counteract the gravity at the equilibrium, whereas $f_e\in\mathcal{F}_{\rm int.for.}^{\rm eq}$ means that the vehicle generates more total thrust than what is strictly necessary for compensating gravity, thus having some wasted internal force. Moreover, there exists a unique orientation for each propeller in an MAV such that when all the propellers assume such orientation the absence of internal force is guaranteed at hovering, proven in Lemma 1 in Sec.~\ref{LemmaProof}~\cite{Ali2025}. The following proposition establishes the main result of this section.
\begin{prop}[Equilibrium Set of the System]
\label{Prop1}
Consider any pose for the base $B_1$ denoted by $C^0_{1e}$. Then the following statements hold true and are equivalent for the system~\eqref{statespace}:
\begin{enumerate}
    \item For any base pose $C^0_{1e}\in SE(3)$, there exists an equilibrium for PJD$_2$ in which the joint equilibrium configuration ${\color{black}{q_{pe}}}^{[2]}$ satisfies $C^0_{j}(C^0_{1e},q^{[2]}_{je})e_3= e_3$ $\forall j \in \{2, \dots, 5\}$ and {\color{black}{ a thrust magnitudes vector $f^{[2]}_{e}\in\mathcal{F}_{\rm min}^{\rm eq}$ is produced whose elements $f^{[2]}_{je}={g_r(\frac{1}{4}m_b+m_j)}$ are constants $\forall C^0_{1e}$ with $k_i$ chosen by $\sum_{j=2}^{5}k_i{(\frac{1}{4}m_b+m_j)}=0$.}}
    \item For any base pose $C^0_{1e}$, PJD$_2$ has no internal force at any equilibrium ${\color{black}{q_{pe}}}^{[2]}$, by Definition \ref{Def1}, as a corollary of $f^{[2]}_{e}\in\mathcal{F}_{\rm min}^{\rm eq}$ (Statement 1).
    \item When $\tau_{dj} \neq 0 $, PJD$_1$ has no joint equilibrium ${\color{black}{q_{pe}}}^{[1]}$ in which each propeller produces a minimal thrust: $C^0_{j}(C^0_{1e},q^{[1]}_{je})e_3\neq e_3,\, \forall \tau_{dj} \neq 0 $.
    \item When $\tau_{dj} \neq 0 $, it follows from Statement 3, and Lemma \ref{Lemma1} in~\cite{Ali2025}, that PJD$_1$ has an internal force $f^{[1]}_{e}\in\mathcal{F}_{\rm int.for.}^{\rm eq}$. {\color{black}{The amount of such force can be reduced by changing parameters $k_i$,$\,c$ and $m_j$.}} 
\end{enumerate} 
\end{prop}
\begin{proof}
    See Sec.~\ref{proofProp1} of~\cite{Ali2025}.
\end{proof}
The state equilibria set for PJD$_i$ is denoted by $\mathbb{D}^{[i]}_{x_e}$ whereas $\mathbb{D}^{[i]}_{f_e}$ signifies their associated thrust equilibrium sets. {\color{black}{The explicit relations are given in Sec. \ref{EquSets} of~\cite{Ali2025}}}.
    
We conclude the section remarking that the perfect vertical alignment of propellers in any PJD$_2$ equilibrium is possible thanks to each spring torque $\tau_{sje}$ absorbing the residual drag $\tau_{dj}$ affecting the joint variable $q_{j1}$. This allows any joint equilibrium ${\color{black}{q_{pe}}}^{[2]}$ to be dictated only by the gravity acting on the CoMs of the pendulum-like links B$_2$-B$_4$, given that the pivots, i.e., the joints, are passive. 

\begin{remark}
{\color{black}{The design using PJD$_1$ constitutes still an excellent compromise between mechanical simplicity and energy efficiency, with the ability to outperform the state of the art designs when it comes to the amount of internal force needed at the the equilibria, as demonstrated in Sec. \ref{sec:Sim}. Hence, PJD$_2$ is not a structural necessity for the concept utility, in fact, for typical values of the aerodynamic and kinematic parameters, the difference in the amount of internal forces generated in PJD$_1$ design remains significantly small compared to the ideal PJD$_2$ case, as observed in the simulations of Sec.~\ref{sec:Sim}.}} This supports the viability of the concept from a mechanical perspective by using the simpler design PJD$_1$, if one does not want to resort to the particular joint design in PJD$_2$.
\end{remark}
\section{Stabilization of the Equilibria}
\label{sec:Cont}
\begin{figure*}
    \centering
    \includegraphics[width=1\linewidth]{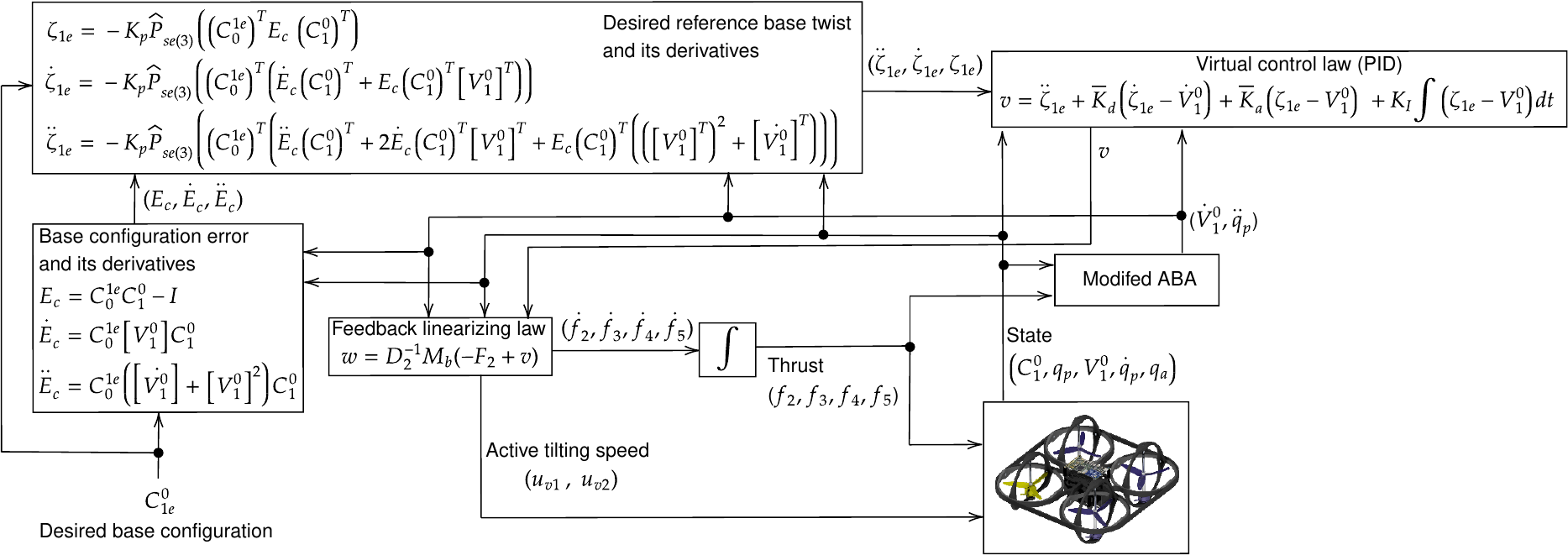}
    \caption{{\color{black}{Block diagram of the  control scheme. It takes a desired base pose $C^{0}_{1e}\in SE(3)$ and converts it into a reference base twist $\zeta_{1e}(t)$ through configuration error feedback $E_c$. Dynamic FBL is applied to the twist dynamics, which are stabilized using a PID controller. The accelerations are efficiently computed via the modified Articulated-Body Inertia Algorithm (ABA) \cite{lieahmed}, illustrated in Sec. \ref{ABASection} of~\cite{Ali2025}.}}} 
    \label{fig:controlBlkDiagram}
\end{figure*}

{\color{black}{Following the characterization of the Equilibria and Internal-Force properties in Proposition~\ref{Prop1}, this section presents the solution of the problem of whether every equilibrium point in the equilibria set $\mathbb{D}^{[i]}_{x_e}$ can be made closed-loop asymptotically stable, at least locally, by a regular state feedback law, hence establishing the Stabilizability property}}. The core findings are stated in Theorem~\ref{Theorm1} and Proposition~\ref{Prop2}. {\color{black}{The proposed control architecture is shown in Fig.~\ref{fig:controlBlkDiagram}.}}

The analysis departs by noticing that there exists no static state feedback control law rendering the closed-loop system $\Sigma_1$ in~\eqref{statespace} locally asymptotically stable (LAS) in an open neighborhood $\mathbb{U}_{{x}_e} \subseteq \mathbb{X}$ of an equilibrium ${x}_e\in \mathbb{D}^{[i]}_{x_e}$, since it turns out that $\Sigma_1$ does not possess a well-defined vector relative degree at any $x_e\in \mathbb{D}^{[i]}_{x_e}$~~\cite{isidori1999nonlinear}.{\color{black}{ However, the goal of stabilizing this closed-loop system at $x_e$ remains attainable by a dynamic state feedback obtained from the dynamic extension algorithm~\cite{DEA}, as illustrated in Theorem 1.}}
\begin{theorem}[Feedback Linearization Compensator and Zero Dynamics  (inner loop)]
   Let the original system $\Sigma_1$~\eqref{statespace} be decomposed into 
    $\Sigma_2$ and $\Sigma_3$, respectively given by $\Sigma_3:\dot C_1^0=[V_1^0] \,C_1^0$ and this $\Sigma_2$
\begin{equation}
\resizebox{0.91\columnwidth}{!}{$
\Sigma_2:\left\{
\begin{aligned}
         \dot{x}_2 =& {\color{black}{F_2(x)+G_2(x)u}}\\=&\left(\begin{array}{c}
          {\color{black}{\dot{q}_{p}}}\\
          M^{-1}(-h-g+\bar{\tau}_{f})\\
          0_{2\times1}
     \end{array}\right)+\left(\begin{array}{c}
          0_{6\times6}  \\
          M^{-1} \left(\begin{array}{cc}
                D_{b} & 0_{6\times2} \\
                 D_{t} & 0_{6\times3}\\
          \end{array}\right)\\
          \begin{array}{cc}
          0_{2\times4} & I_{2\times2}
          \end{array}
     \end{array}\right) u
\end{aligned}
\right.$}
     \label{x2}
\end{equation}
   Then, these statements are true:   
\begin{enumerate}
    \item  $\Sigma_2$ is dynamically input/output feedback linearizable w.r.t the I/O pair $(u,h_y)$, where the output function $h_y(t): t \rightarrow \mathbb{R}^{6}$ is $h_y(t)={\color{black}{V_1^0}}$, by a dynamic extension~\cite{DEA} given by this input precompensator 
\begin{equation}
  \resizebox{0.83\columnwidth}{!}{$  z(t)=col(z_i):=f(t) \in \mathbb{R}^4,  \dot{z}=\bar{w} \in \mathbb{R}^4, i=2,\ldots,5.
    $}\label{precompensator}
\end{equation}The extended state is $\bar{x}=({\color{black}{q_{p}}},{\color{black}{V_1^0,\dot{q}_{p}}},{\color{black}{q_a}}, z)\in \bar{\mathbb{X}}\equiv \mathbb{T}^8\times \mathbb{R}^{16}$ whereas the linearizing state feedback on the extended system, whose I/O pair is $(w,h_y)$, is given by
\begin{equation}
    {w}= \left(\begin{smallmatrix}
        \Bar{w}\\ u_{v_1} \\u_{v_2}
    \end{smallmatrix}\right)=D_2^{-1}M_b(-E_2+v) \label{lineariningW}
\end{equation}
where $v\in \mathbb{R}^{6}$ is the virtual input, and $E_2$, $D_2$ and $M_b$ are calculated from, with $j=2,3,4$ 
{\color{black}{\begin{equation}
\resizebox{0.95\columnwidth}{!}{$
\begin{aligned}
D_2 &=
\begin{bmatrix}
N_j\operatorname{Ad}^T_{(C^0_{f_j})^{-1}}\bar f_j
& \cdots &
\operatorname{Ad}^T_{(C^0_{f_5})^{-1}}\bar f_5
& \bar D_2
\end{bmatrix},
\\
\bar D_2 &=
\begin{bmatrix}
-\operatorname{ad}^T_{S_{51}}
\operatorname{Ad}^T_{(C^0_{f_5})^{-1}}\bar f_5 z_5
&
-\operatorname{ad}^T_{S_{52}}
\operatorname{Ad}^T_{(C^0_{f_5})^{-1}}\bar f_5 z_5
\end{bmatrix},
\\
E_2 &=
M_b^{-1}\Big(
-\dot M_{bp}\big(L_1+M_{pp}^{-1}W^0_{\mathrm{app},p}\big)
-M_{bp}L_2
\\
&\qquad\qquad
+\big(\dot M_{bp}M_{pp}^{-1}M_{bp}^T-\dot M_{bb}\big)\dot V^0_1
-\dot h_b-\dot g_b+\bar W
\Big),
\\
\bar W &=
-\sum_{j=2}^{4}
B_j\operatorname{ad}^T_{V^0_{j2}}W^0_{\mathrm{app},B_j}
-\operatorname{ad}^T_{V^0_1}W^0_{\mathrm{app},B_5},
\\
L_1 &:=
M_{pp}^{-1}\left(\tau_f-h_t-g_p\right),
\\
L_2 &:=
M_{pp}^{-1}
\left[
\left(\dot\tau_f-\dot h_p-\dot g_p\right)
-\left(\dot M_{bp}^T\dot V^0_1+\dot M_{pp}\ddot q_p\right)
\right],
\\
M_b &:=
M_{bb}-M_{bp}M_{pp}^{-1}M_{bp}^T,
\\
N_j &:=
I_6
-
M^0_{B_j}S_j
\left(S_j^TM^0_{B_j}S_j\right)^{-1}
S_j^T.
\end{aligned}$}
\end{equation}}}
This dynamic FBL is well-defined in an open neighborhood $\mathbb{U}_{\bar{x}_e}\subseteq \bar{\mathbb{X}}$ of $\bar{x}_e \in \mathbb{D}^{[i]}_{x_e} \cup \mathbb{D}^{[i]}_{f_e}$ defined as
\begin{equation}
       \mathbb{U}_{\bar{x}_e}=\Big\{\bar{x}(t)\in \bar{\mathbb{X}}
   \,\Bigr\rvert \, \operatorname{det}\big(D_2(\bar{x})\big)\neq0
\Big\}
 \label{SingularitySet}
\end{equation}
    \item For any base pose ${\color{black}{C_{1e}^0}}\in SE(3)$, The original system $\Sigma_1$, extended with the precompensator~\eqref{precompensator}, has locally, in $\mathbb{U}_{\bar{x}_e}\times SE(3)$, asymptotically stable (LAS) zero dynamics w.r.t the output $h_y$, iff $\,\forall  \bar{x} \in \mathbb{Z}_x-\{\bar{x}_e\}$
    \begin{equation}
         \|\tilde{f}\|< \frac{1}{\|D_3({\color{black}{q_p}}-{\color{black}{q_{pe}}})\|} \lambda_{\text{min}}(D_f) \|{\color{black}{\dot{q}_p}}\| 
         \label{FinalBoundsZD}
\end{equation} 
where $\tilde{f}=col({f}_i)\,, i=2,3,4$,   $\mathbb{Z}_x$ and $D_3$ are respectively given in \eqref{ZeroDOriginal} and \eqref{thrustZD} in~\cite{Ali2025}. $D_f>0 \in \mathbb{R}^{6\times6}$ contains the constant passive joints friction coefficients.
\end{enumerate}  
\label{Theorm1}
\end{theorem}
\begin{proof} See Sec.~\ref{proofTheorem1} in~\cite{Ali2025}.\end{proof}
This means that, as long as the extended system trajectories remain in $\mathbb{U}_{\bar{x}_e}\times SE(3)$ where the decoupling holds, any base pose $C^0_{1e}$ of the original system $\Sigma_1$~\eqref{statespace} can be made locally asymptotically stable by a dynamic state feedback. Proposition \ref{Prop2} constructs the the virtual controller that stabilizes the resulting linearized twist dynamics. 
\begin{prop}[Asymptotic Stabilization of Equilibria (outer loop)]
\label{Prop2}
Consider the virtual control input $v$ given by
\begin{equation}
v=\ddot{\zeta}_{1e}+\bar{K}_d(\dot{\zeta}_{1e}-\dot{V}^0_1)+\bar{K}_a({\zeta}_{1e}-V^0_1) \label{Virtual}
\end{equation}
where $\zeta_{1e}$, $\dot{\zeta}_{1e}$, $\ddot{\zeta}_{1e}$ $\in \mathbb{R}^{6}$ are respectively given in~\eqref{zeta_1e} and~\eqref{timeDerizeta1e}~\cite{Ali2025} and shown in Fig. \ref{fig:controlBlkDiagram}. $\bar{K}_d$ and $\bar{K}_a$ $\in \mathbb{R}^{6\times 6}$ are positive definite diagonal gain matrices chosen such that to make $(A-BK)$ Hurwitz where 
\begin{equation}\resizebox{0.95\columnwidth}{!}{$
    A=\left(\begin{array}{c}
                0_{6\times6}\quad I_{6\times6}   \\
                0_{6\times6} \quad 0_{6\times6}\\
\end{array}\right),\, B=\left(\begin{array}{c}
                0_{6\times6}   \\
               I_{6\times6}\\
\end{array}\right),\, K= \left(\begin{array}{c}
                \bar{K}_a\quad \bar{K}_d 
\end{array}\right)$}
\end{equation}Given the precompensator in~\eqref{precompensator} and the linearizing control law in~\eqref{lineariningW}, substituting $v$ into~\eqref{lineariningW} asymptotically stabilizes the extended system $\Sigma_1$ with~\eqref{precompensator} in the open neighborhood $\mathbb{U}_{\bar{x}_e}\times SE(3)$.
\end{prop}
\begin{proof} See Sec.~\ref{proofProp2} in~\cite{Ali2025}. \end{proof}
{\color{black}{We note that the system \eqref{statespace} evolves on the product Lie group 
$SE(3)\times\mathbb{T}^8$ due to the internal joint dynamics. This motivates our use of the normal-form approach~\cite{isidori2013nonlinear}, as it facilitates the study of the associated zero dynamics. Although the proposed 
controller, shown in Fig. \ref{fig:controlBlkDiagram}, employs tools often considered coordinate-dependent, the Lie-group 
formulation guarantees that the control laws are coordinate-invariant, i.e., their validity does not depend on a specific local parameterization of the configuration $SE(3)\times\mathbb{T}^8$.}}

{\color{black}{The controller implementation requires measurements of the base pose and twist, passive-joint positions and velocities, and active-joint positions. These quantities can be obtained using standard sensing and estimation pipelines for aerial robots, for example by fusing IMU measurements with motion-capture, odometry, or GPS data through a UKF, together with compact joint encoders and velocity observers. The computed thrust commands are then tracked by low-level ESC speed loops governing the propellers' brushless DC motors, while the active-link tilting commands are realized by the corresponding servo motors drivers. In practice, motor and servo bandwidths, thrust saturation, and rate limits constrain the admissible maneuver speed. Nevertheless, in the simulations reported in Sec.~\ref{sec:Sim}, the required thrusts and tilting rates remain close to their hovering values and are compatible with commercially available propellers and servomotors, such as Dynamixel servo motors.}}
\section{Simulation Results}
\label{sec:Sim}
\begin{figure*}
   \centering
\begin{subfigure}{0.475\linewidth}
\includegraphics[width=1\columnwidth]{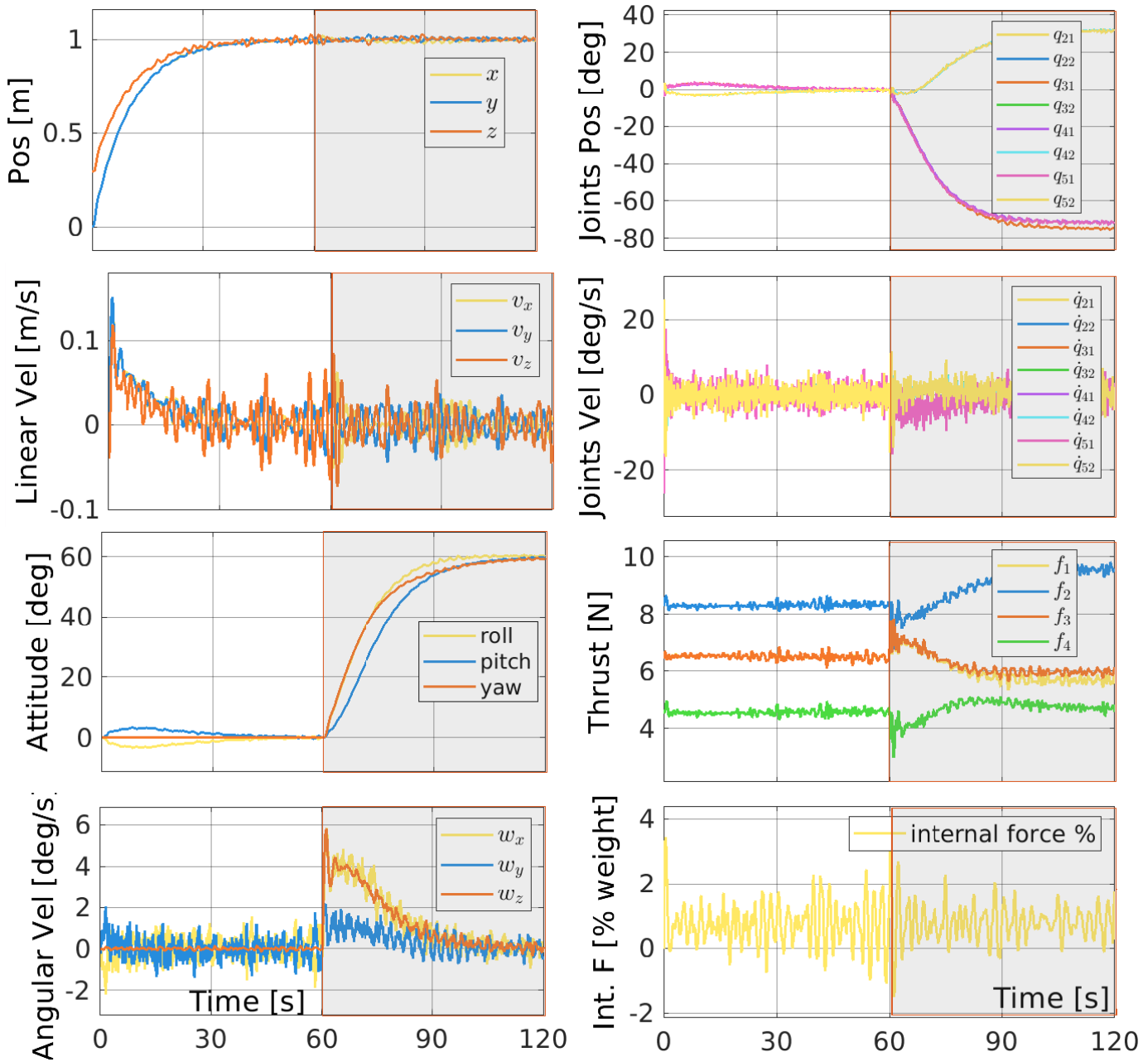}
\caption{PJD$_1$}
\label{Sim_PJD1}
\end{subfigure}
\begin{subfigure}{0.475\linewidth}
\includegraphics[width=1\columnwidth]{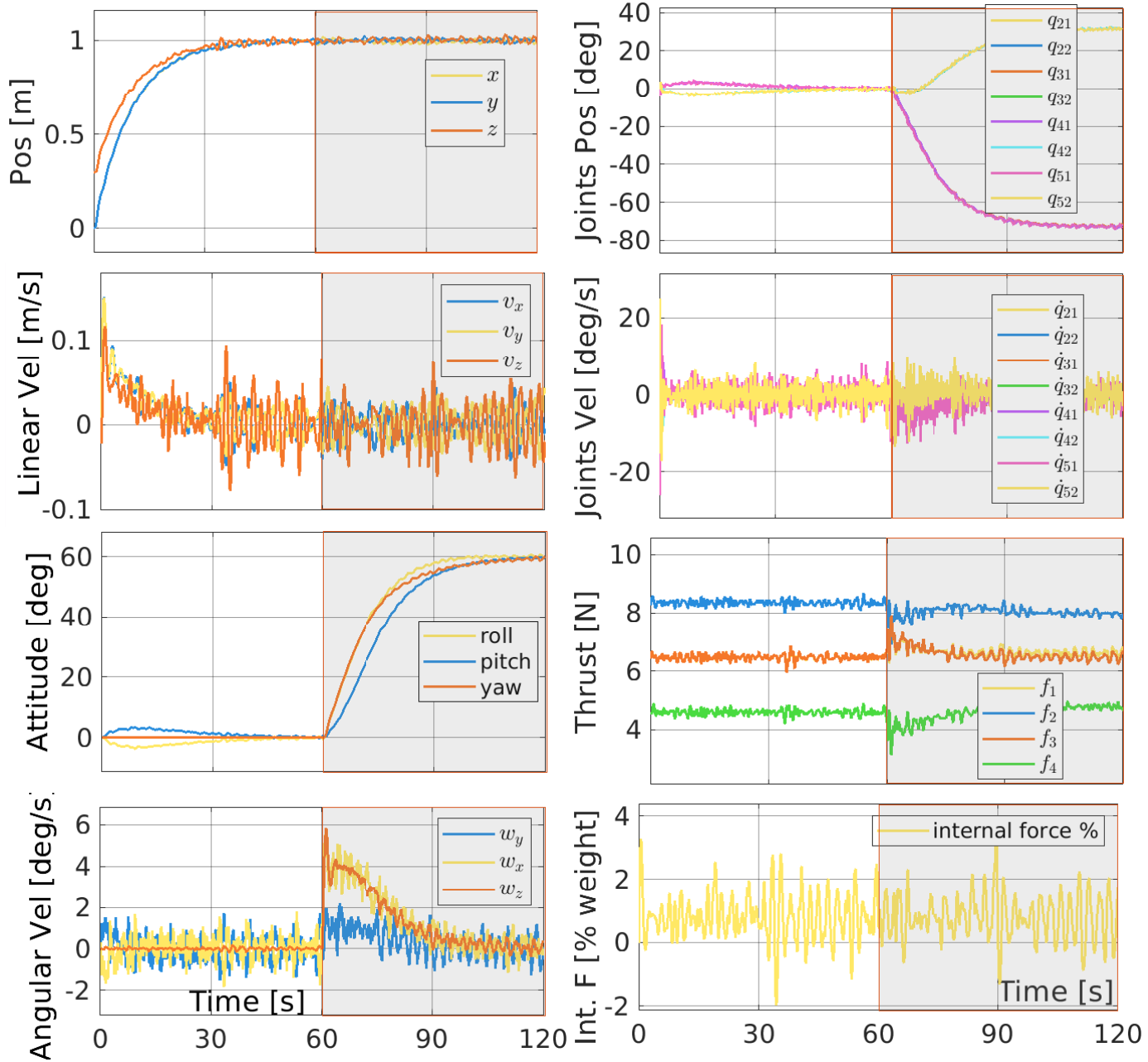}
\caption{PJD$_2$}
\label{Sim_PJD2}
\end{subfigure}
\setlength{\belowcaptionskip}{-8pt}
    \caption{{\color{black}{Evolutions of the state and input for (a) PJD$_1$ and (b) PJD$_2$ designs. The simulation has two phases: First (0-60s) is the position step with the initial attitude held constant. Second (60-120s, gray-shaded), the attitude step with the position at $t=60$s held constant.
    }}}
    \label{Sim1}
\end{figure*}

{\color{black}
This section reports high-fidelity Gazebo simulations, implemented via the GenoM3 framework, validating the proposed controller beyond the nominal model and under more realistic conditions such as presence of parametric uncertainty and input noise. {\color{black}{We also compare the performance of the PenduMAV, in a 6-DoF stabilization maneuver, against 6 different fully-actuated designs from the literature}}. For both designs PJD$_1$ and PJD$_2$, the controller~\eqref{lineariningW}, \eqref{Virtual} with the dynamic extension~\eqref{precompensator}, shown in Fig. \ref{fig:controlBlkDiagram}, achieves independent stabilization of the MAV attitude and position. Two indicative desired base poses are considered: an attitude command of a $60^\circ$ step in roll, pitch and yaw at fixed position, and a $1$~[m] position step along the inertial $(x,y,z)$ with constant orientation. The resulting trajectories are shown in Fig.~\ref{Sim1}.} {\color{black}{Simulation videos and plots for several different maneuvers are available at \url{https://www.youtube.com/playlist?list=PL4N8pJgvqASQX6AWEpg3NCZ6QdGBPfbXq}}. 
Most notably, a simulation with $270$-degree roll turn is included (see also Fig.~\ref{fig:3DComplete} and~\ref{Sim270}~\cite{Ali2025}) which demonstrates full omnidirectional capabilities.  
As it can be noted from the simulations, the vehicle achieves a wide range of maneuvers while generating only positive thrust, potentially reducing the complexity associated with the use of bidirectional propellers. 
The state and input evolutions exhibit the natural real-world-like oscillations due to the imperfections introduced by parametric uncertainty and actuator noise from the emulated hardware.}

{\color{black}{The simulations use $m_b=1,m_j=0.479, m_5=0.12$~kg, with $B_j$'s inertias (kg$\cdot$m$^2$) $I_{px}=I_{py}=2.18\times10^{-3}$, $I_{pz}=1.17\times10^{-6}$, and $B_1$'s inertias $I_{bx}=I_{by}=8.33\times10^{-3}$ and $I_{bz}=1.67\times10^{-2}$. The distance (m) between adjacent joints is $a=0.2$ whereas from each joint to the corresponding link CoM is $c=0.16$. Friction coefficients are $b_{f_{j1}}=b_{f_{j2}}=0.05$~N$\cdot$m$\cdot$s/rad, and the $k_j$ ratio is $0.01$. These values are each perturbed by $+10\%$. 
    The gains $\bar{K}d=\operatorname{diag}(5.55)$ and $\bar{K}a=\operatorname{diag}(14.94)$ are obtained via LQR applied to the system in Prop~\ref{Prop2}, using weighting matrices ${\color{black}{W_x}}=\operatorname{diag}(0.0001)\in\mathbb{R}^{12\times12}$ and ${\color{black}{W_u}}=\operatorname{diag}(0.0001)\in\mathbb{R}^{6\times6}$. The gain $K_I=\operatorname{diag}(30)$ is tuned empirically to eliminate steady-state error. The gain $K_p=\operatorname{diag}(0.1)$ in~\eqref{zeta_1e}~\cite{Ali2025} regulates the convergence rate of $C^0_1$ to $C^0_{1e}$, hence the evolution speed of the reference trajectory $\zeta_{1e}$. For fixed $\bar{K}_d$ and $\bar{K}_a$, increasing $K_p$ leads to larger transient errors, requiring higher thrust and faster tilting for $q_a$.}}   

\begin{figure}
    \centering
\includegraphics[width=1\columnwidth]{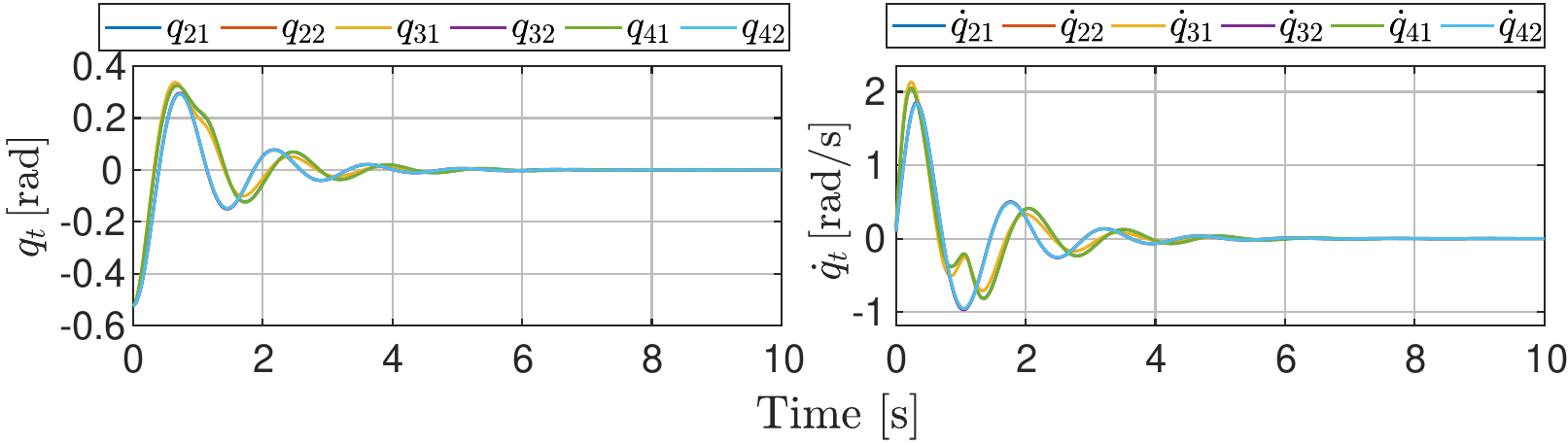}

    \caption{zero dynamics evolution showing convergence.}
    \label{Case1_Sim_ZD}
\end{figure}

To illustrate the stability of the zero dynamics, as proven in Sec.~\ref{proofTheorem1} of ~\cite{Ali2025}, we simulated its evolution under nominal conditions with initial values $C^0_1(0) = I_{4 \times 4},\,{\color{black}{q_p}}(0) = -30 ^\circ \times \boldsymbol{1}_{6 \times 1} \,$, $\dot{{\color{black}{q_p}}}(0) = 0.1\times \boldsymbol{1}_{6 \times 1}$ rad/s. As shown in Fig.~\ref{Case1_Sim_ZD}, the $q_p$ and $\dot q_p$ converge to zero, confirming convergence to the equilibrium $q_{pe}$ corresponding to that base pose $C^0_{1e}(0)$.

\begin{figure}
    \centering
\includegraphics[width=0.65\columnwidth]{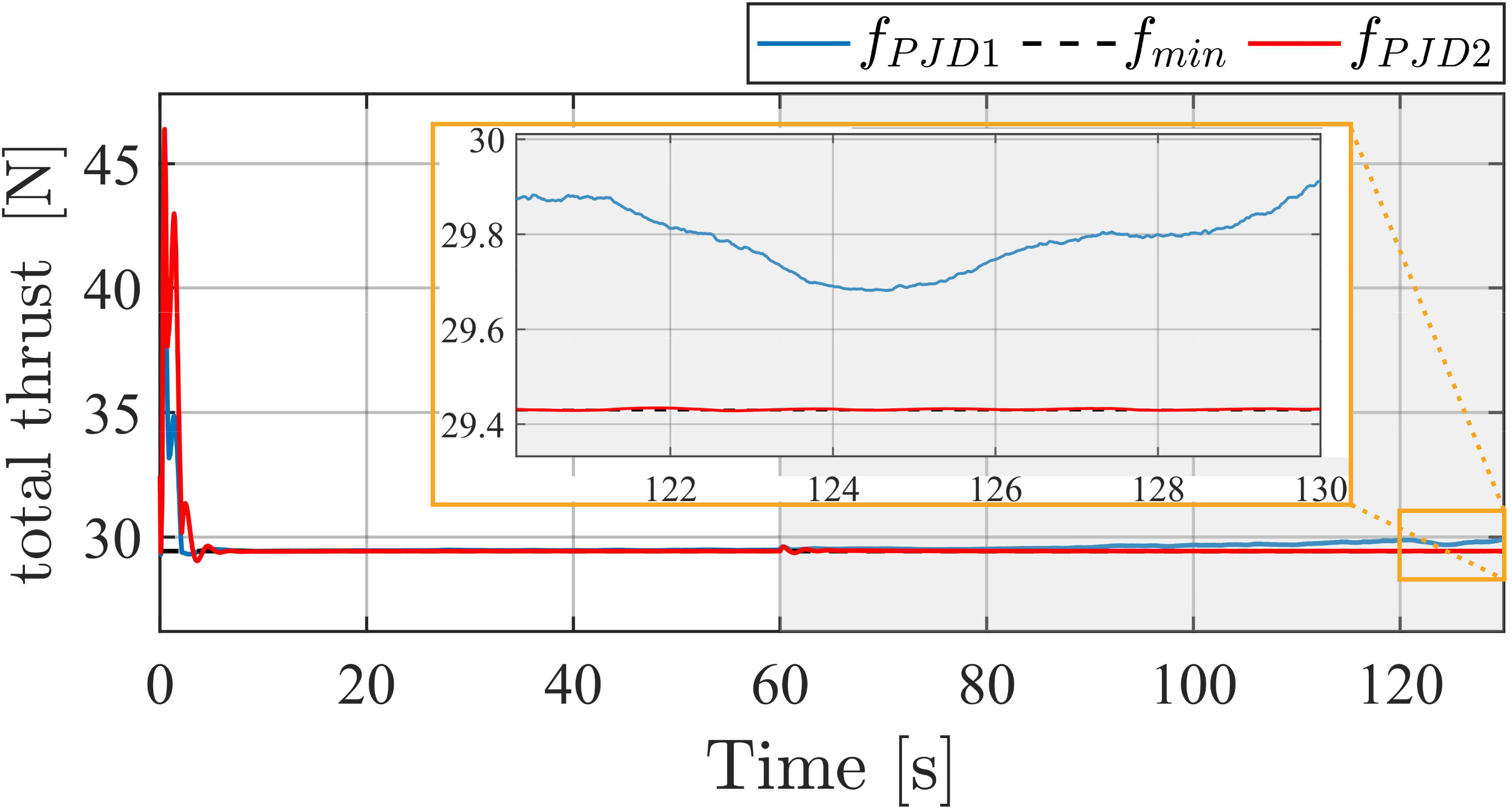}
    \caption{Comparison of the total thrust between PJD$_1$ and PJD$_2$ when stabilizing a $60^\circ$ pitch attitude step. In hovering, $\sum f^{[1]}_{e}$ is 2\% higher than $ \sum f^{[2]}_{e}$.}
    \label{ComparsionThrust}
\end{figure}

We conclude by comparing PJD$_1$ and PJD$_2$ in terms of the total equilibrium thrust for a scenario where a step command of $60^\circ$ in pitch is given. As plotted in Fig.~\ref{ComparsionThrust}, the equilibrium thrust of PJD$_1$ satisfies $\sum^{5}_{j=2}f^{[1]}_{je}=1.02 \sum^{5}_{j=2}f^{[2]}_{je}$ with $f^{[2]}_{e}=f_{min}\in \mathcal{F}_{\rm min}^{\rm eq}$, i.e. a 2$\%$ increase over the minimal thrust $f^{[2]}_{e}$ in PJD$_2$, since the desired pose $C^0_{1e}$ requires residual torque $\tau_{dj}\neq 0$, leading to $q_{pe}^{[1]}\neq q_{pe}^{[2]}$, i.e not pointing upward. This confirms Prop~\ref{Prop1}.
\begin{figure}
    \centering
    \includegraphics[width=1\columnwidth]{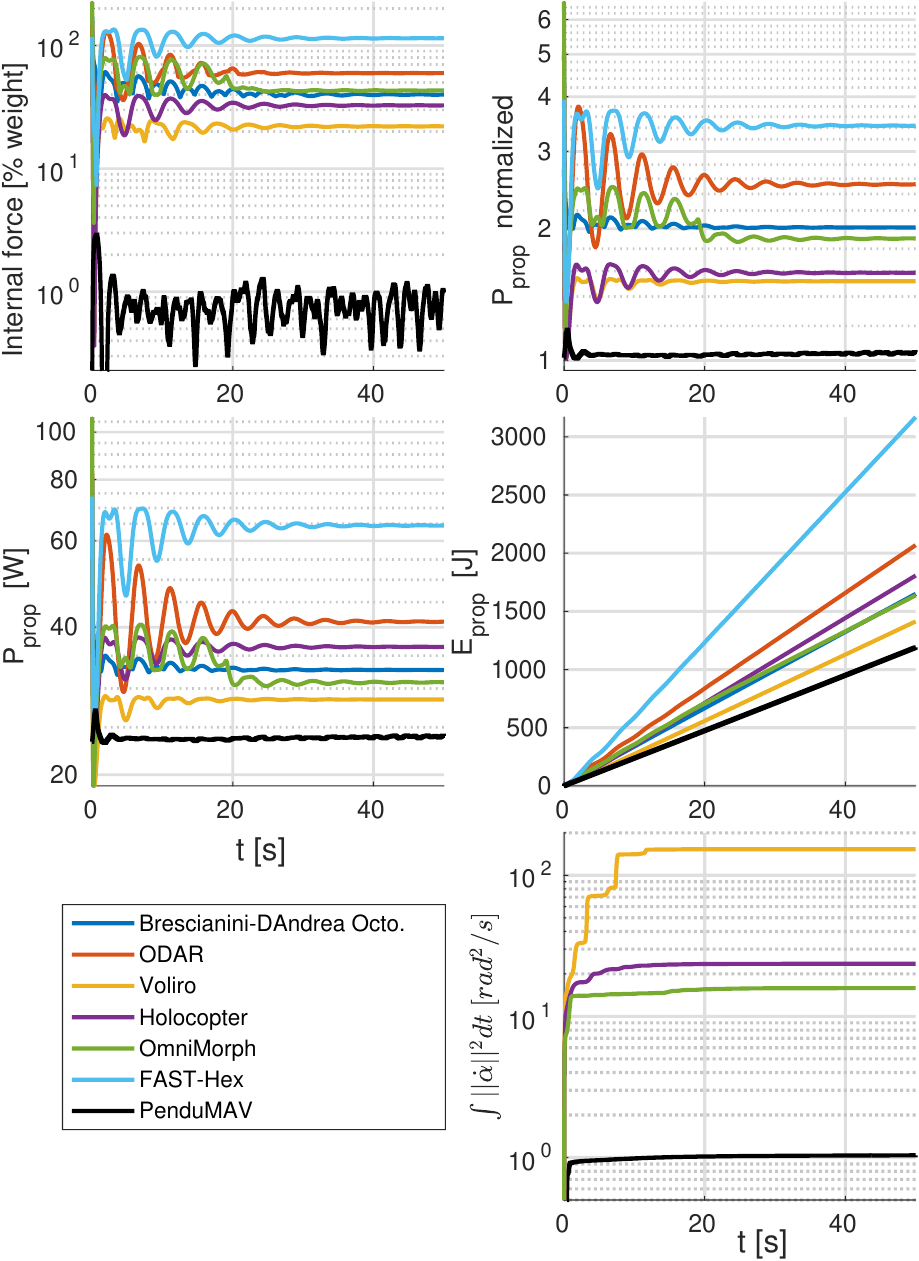}
    \caption{{\color{black}{Comparison among fully actuated MAVs stabilizing the same setpoint. Under the common simulation benchmark, the PenduMAV achieves the most favorable values for all reported metrics.}}}
    \label{fig:vehcomp}
\end{figure}

\begin{figure*}[t]
    \centering

    \begin{minipage}[t]{0.49\textwidth}
        \centering
        \vspace{0pt}
        \begin{subfigure}[t]{1\textwidth}
            \centering
            \includegraphics[width=\textwidth]{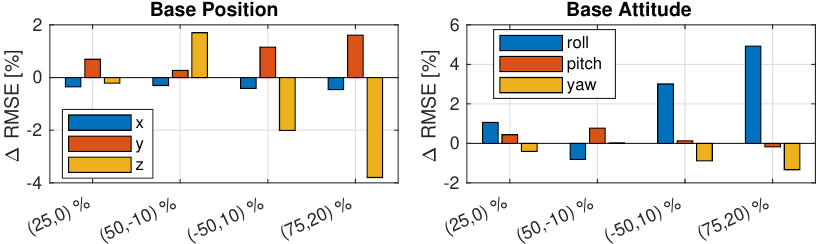}
            \caption{}
            \label{fig:rmse_uncert_inertia_mass_b5}
        \end{subfigure}

        \begin{subfigure}[t]{1\textwidth}
            \centering
            \includegraphics[width=\textwidth]{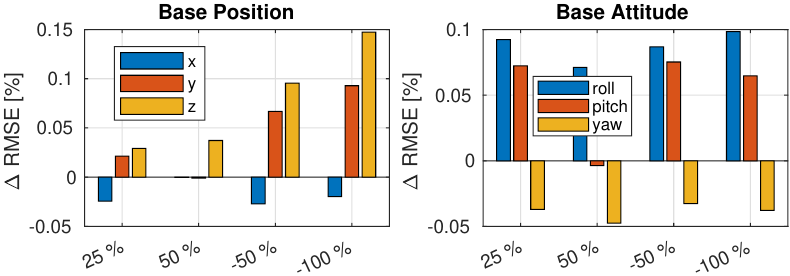}
            \caption{}
            \label{fig:rmse_uncert_friction}
        \end{subfigure}

        \begin{subfigure}[t]{\textwidth}
            \centering
            \includegraphics[width=\textwidth]{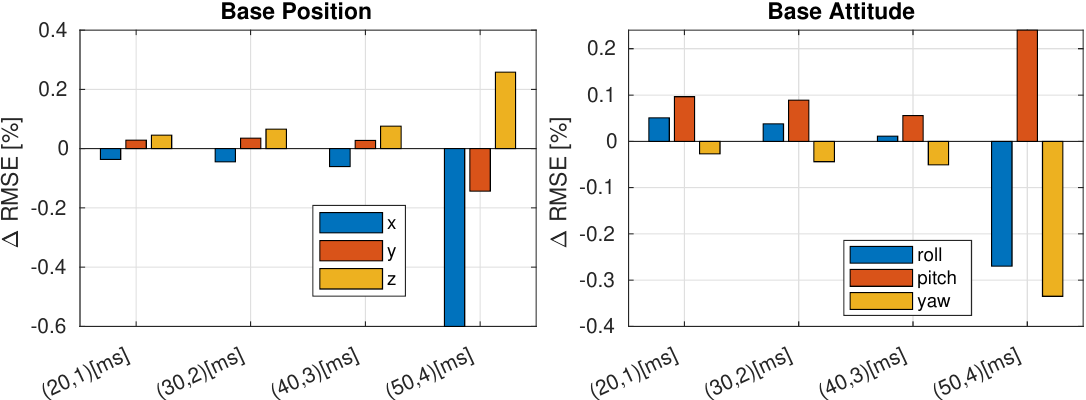}
            \caption{}
            \label{fig:rmse_motor_delay}
        \end{subfigure}

        \begin{subfigure}[t]{\textwidth}
            \centering
            \includegraphics[width=\textwidth]{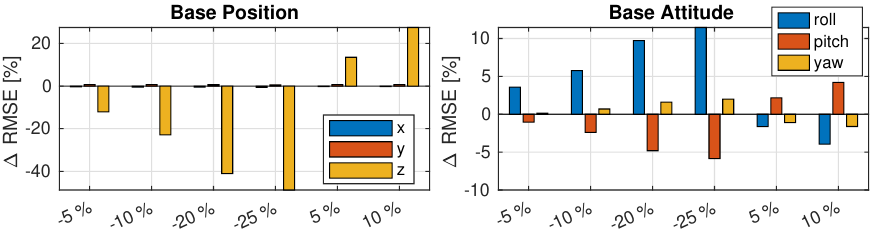}
            \caption{}
            \label{fig:rmse_uncert_cf}
        \end{subfigure}
    \end{minipage}
    \hfill
    \begin{minipage}[t]{0.49\textwidth}
        \centering
        \vspace{0pt}

        \begin{subfigure}[t]{\textwidth}
            \centering
            \includegraphics[
                width=\textwidth,
                height=0.86\textheight,
                keepaspectratio
            ]{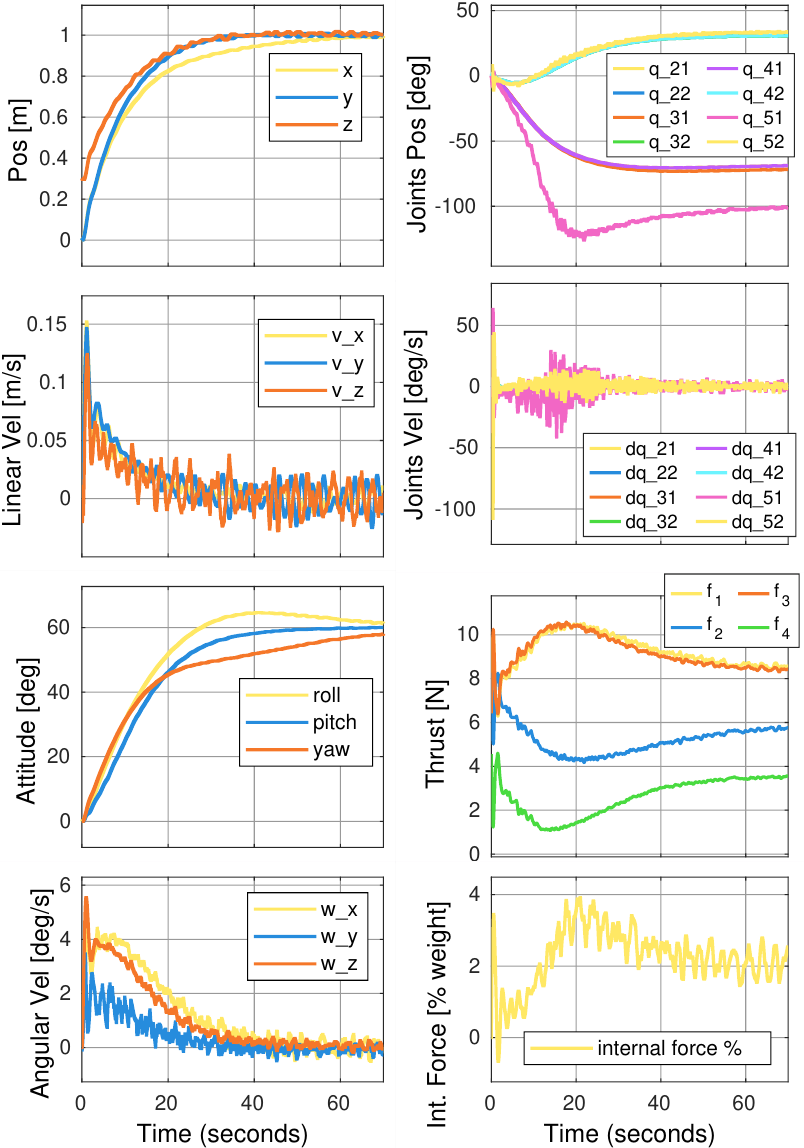}
            \caption{}
            \label{fig:rmse_backlash}
        \end{subfigure}
    \end{minipage}

    \caption{
        \color{black}{RMSE percentage change over the corresponding nominal RMSE. While All cases were stable, the transit performance were affected by the perturbation. The comparison is under different uncertainty and disturbance cases:
        (a) (inertia, mass) $\%$ mismatch of the actively-tilting body $B_5$, 
        (b) friction uncertainty,
        (c) different pairs of (time constant, delay) values,
        (d) thrust coefficient uncertainty, and
        (e) $5^\circ$ backlash effect. 
    }}
    \label{fig:rmse_uncertainty_backlash}
\end{figure*}

{\color{black}{To provide a quantitative assessment of how PenduMAV compares with popular state-of-the-art platforms, we perform a common stabilization simulation where the setpoint is $(x,y,z, \text{roll},\text{pitch},\text{yaw})=(1,1,1,60^\circ,60^\circ,60^\circ)$, in which the PenduMAV performance is evaluated against that of these six different fully-actuated multirotors:
Brescianini-D'Andrea Octorotor \cite{8reversible}, ODAR \cite{ODAR8inputs}, Voliro \cite{Voliro12inputs}, Holocopter \cite{ryll2014novel}, OmniMorph \cite{aboudorra2024modelling},  and
FAST-Hex \cite{FastHex} with bidirectional propellers. All vehicles are simulated using the same rigid-body and aerodynamics parameters, controller design and tuning, actuator limits, and simulation settings. Therefore, the comparison reflects solely the effect of how the same desired wrench is allocated to these vehicles' different actuation structures. The results are reported in Fig.~\ref{fig:vehcomp}.
The comparison is based on these metrics: internal-force
percentage \(I=\left(\frac{\sum_i |f_i|}{\text{weight}}-1\right)\times\%100\), instantaneous
propulsive power
\(P_{\mathrm{prop}}=\sum_i c_t\left(|f_i|/c_f\right)^{3/2}\),
accumulated propulsive energy
\(E_{\mathrm{prop}}=\int_0^T P_{\mathrm{prop}}(t)\,dt\), normalized
propulsive power
\(\bar{P}=\sum_i |f_i|^{3/2}/\left(N(\text{weight}/N)^{3/2}\right)\), and cumulative
tilt activity
\(J_{\alpha}=\int_0^T \sum_i \dot{\alpha}_i^2(t)\,dt\). These metrics roughly
quantify, respectively, excess or mutually canceling thrust, rotor power demand, total energy consumed by propellers, thrust-distribution
efficiency among propellers ($\bar{P}=1$ means all propellers produce the same thrust), and the mechanical activity required from the tilting mechanisms.}}

{\color{black}{Furthermore, to assess the sensitivity of the controller to individual perturbations, we carried out a set of simulations in which different sources of model mismatch were injected separately over a wide range of values. The resulting RMSE trends of the pose, computed discretely from \(\mathrm{RMSE}_{p}=\sqrt{\frac{1}{N}\sum_{k=1}^{N}\left\|p(t_k)-p_d\right\|^2}.\) where $p$ is a component of the pose and $p_d$ its setpoint, with respect to the nominal RMSE, are reported in Fig.~\ref{fig:rmse_uncertainty_backlash} for uncertainties in the inertia and mass of $B_5$, passive joints friction, thrust coefficient $c_f$. Moreover, the effect of different motor delays and time constants along with passive-joint backlash on the RMSE in the pose evolution is illustrated respectively in Fig.~\ref{fig:rmse_uncertainty_backlash} (c) and (e). While in all cases the stabilization succeeded, this comparison isolates the contribution of each perturbation source and shows how the stabilization performance might degrade as the mismatch level increases.

The backlash in the six passive joints is modeled as an angular free-play region, centered around $q_0=0$, of half-width \(\delta=2.5^\circ\). For each passive joint, the torque induced by the backlash is \(\tau=-k_c(q-q_0-\delta)-d_c\dot q\) for \(q-q_0>\delta\), and \(\tau=-k_c(q-q_0+\delta)-d_c\dot q\) for \(q-q_0<-\delta\) where \(k_c=0.03~\mathrm{Nm/rad}\), \(d_c=0.03~\mathrm{Nms/rad}\). Thus, inside the backlash gap no additional torque is applied, whereas outside of it, the joint gear contacts the side of that gap and hence faces a small restoring torque plus damping.
}}

{\color{black}{To summarize, as evident from the comparison with representative fully actuated MAVs in Fig.~7 of Sec.~V, the PenduMAV concept is more energy efficient because it avoids expending energy on redundant actuation and on generating excessive thrust. 
Although the proposed concept actively actuates one propeller, the required tilting effort can be significantly lower than in alternative designs with multiple actively tilted propellers. Moreover, by replacing the mass associated with additional actuators with larger propellers, the PenduMAV can generate more lift with the same overall weight, 
improving both payload capacity and endurance. Mechanical reliability is also increased in three important aspects. First, the total thrust is more evenly distributed among the propellers than in the other vehicles, helping prevent overloading of any single propeller. Second, over a wide range of maneuvers tested in simulation, the commanded thrust remains positive, potentially eliminating the need for bidirectional propellers and their associated allocation complexity. Third, because smaller internal forces act on the chassis, the structure can be manufactured to be more lightweight without increasing the risk of vibration or excessive structural loading.}}

\section{Conclusion}
\label{Conclusion:cont}
\begin{figure}[t]
    \centering
    \includegraphics[width=1\columnwidth]{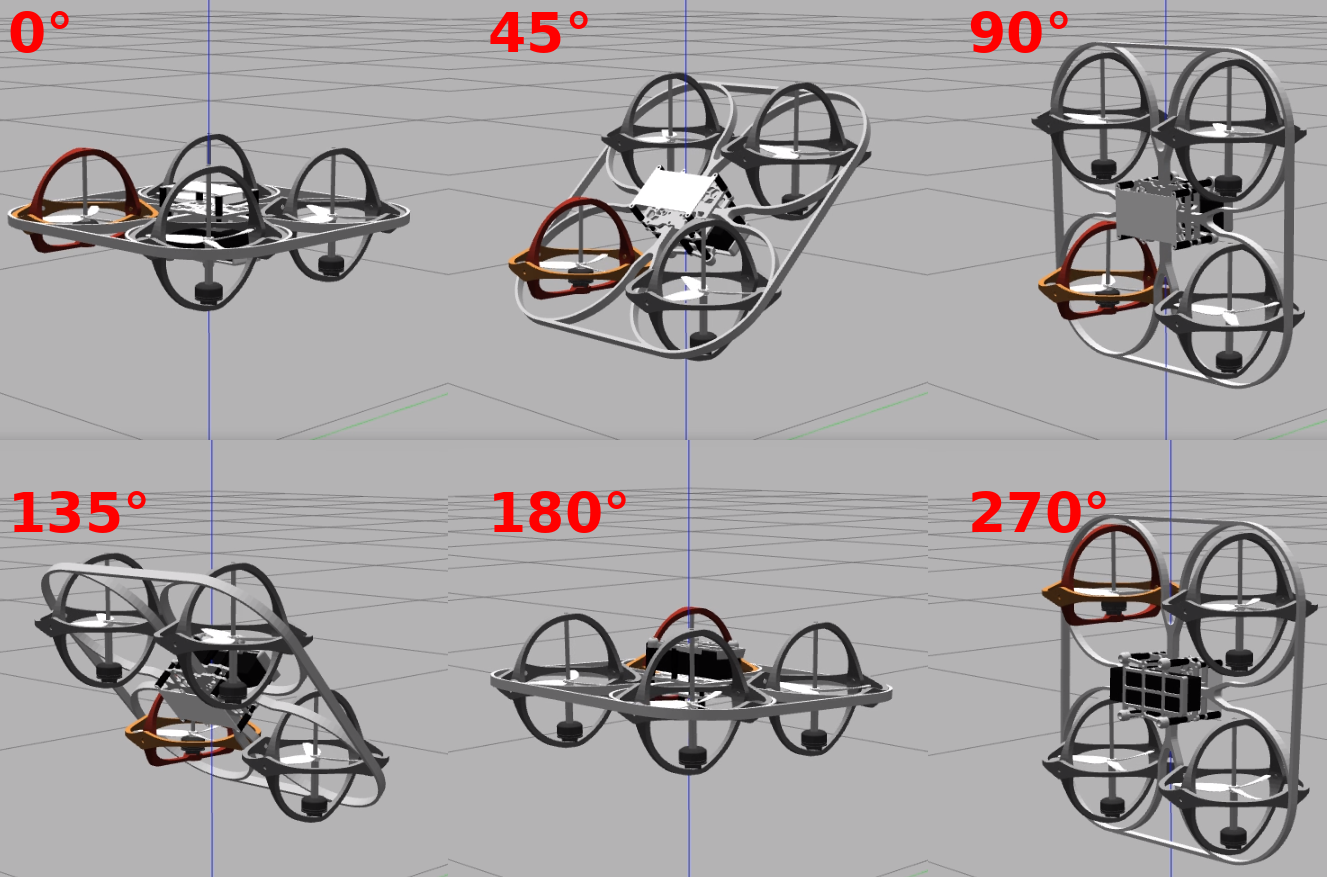}\\
\vspace{1em}
\includegraphics[width=1\columnwidth]{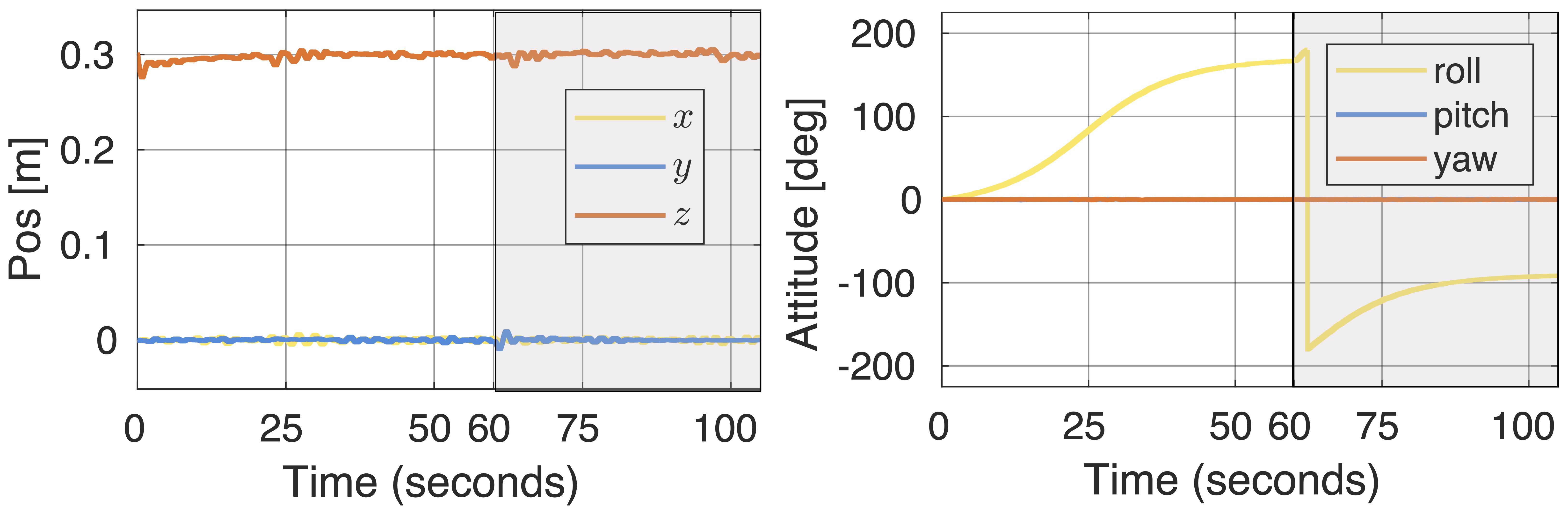}
    \caption{\color{black}{Top: the PenduMAV, simulated with an independent and uncertain/noisy model in Gazebo with the DART physics engine, performing a   270-degree roll maneuver. Video with different maneuvers available at \url{https://www.youtube.com/playlist?list=PL4N8pJgvqASQX6AWEpg3NCZ6QdGBPfbXq}}.\\ Bottom: Position and attitude time evolutions of the two phases.}
\label{fig:3DComplete}
\end{figure}
This work proposed PenduMAV, a novel omnidirectional MAV concept enabling full 6-DoF motion with a minimum number of actuators and no internal forces at equilibrium. The design combines a rigid base 
with 4 propeller-carrying links connected through universal joints, 3 of which are passive whereas 1 is actively driven by 2 servomotors. We proved that the proposed MAV satisfies the Input-Dimension, Equilibria, Internal-Force, and Stabilizability properties. An equilibrium exists for any base pose, and when torsion springs are employed, the total thrust exactly balances the vehicle weight. Even without springs, internal forces can be made arbitrarily small through changing the links masses and lengths. Any base pose can be rendered locally asymptotically stable via a dynamic FBL and backstepping from base pose error on $SE(3)$. {\color{black}{High-fidelity simulations demonstrate decoupled position and attitude control under realistic conditions.}}

 {\color{black}{This design scales favorably, as hovering loads are distributed almost evenly among propellers, avoiding excessive stress on the actively tilting unit. Scaling challenges are therefore those typical of multirotors, such as propeller and battery sizing, rather than limitations specific to the proposed MAV. Future work will focus on realizing, implementing, and testing the mechanical concept.}}

\bibliographystyle{IEEEtran}

\appendix
\label{append}
\subsection{Explicit Expressions of dynamical model}
\label{app:ModelDefinitions}
In this subsection, we collect the definitions and analytical expressions of the variables appearing in the system dynamics \eqref{statespace}. In the following, $j=2,3,4$ and $i=1,2$.
\paragraph{Mass matrix}
The generalized mass matrix of the system is
\[
M =
\begin{pmatrix}
M_{bb} & M_{bp} \\
M_{bp}^T & M_{pp}
\end{pmatrix}.
\]
The base inertia block is
\[
M_{bb} = M^0_{B_1} + \sum_{j=2}^{5} M^0_{B_j},
\]
where $M^0_{B_j}$ denotes the inertia tensor of body $B_j$ expressed in the world frame $\mathcal{F}_w$.
The coupling $M_{bp}$ and joint inertia matrices $M_{pp}$ are respectively defined as
\[
M_{bp} = \sum_{j=2}^{4} M^0_{B_j} S_j, \qquad
M_{pp} = \operatorname{blkdiag}\!\big(S_j^T M^0_{B_j} S_j\big).
\]

\paragraph{Time derivatives}
The time derivative of the inertia tensor of body $B_j$ expressed in $\mathcal{F}_w$ is obtained from
\[
\dot{M}^0_{B_j}
= - M^0_{B_j}  \operatorname{ad}_{V^0_j}
- \operatorname{ad}_{V^0_j}^T M^0_{B_j}  ,
\]
where $\operatorname{ad}_{(\cdot)}$ denotes the Lie-algebra adjoint operator and
\[
V^0_j = V^0_1 + S_j \dot{q}_j ,
\qquad j \in \{2,3,4\}.
\]
For $j=5$, one has
\[
\dot{M}^0_5
= - M^0_5 \operatorname{ad}_{V^0_1}
- \operatorname{ad}_{V^0_1}^T M^0_5 .
\]
The time derivative of the joint screw is given by
\[
\dot{S}_j = \operatorname{ad}_{V^0_j} S_j .
\]

\paragraph{Coriolis and centrifugal terms}
The Coriolis and centrifugal vector is defined as
\[
h = \big( h_b \;\; \operatorname{col}(h_j) \big)^T .
\]
The base component $h_b$ is computed from
\[
h_b
= \dot{M}_{bb} V^0_1
+ \sum_{j=2}^{4}
\big(M^0_{B_j}\dot{S}_j + \dot{M}^0_{B_j} S_j\big)\dot{q}_j ,
\]
while the \(B_j\)'s Coriolis \& centrifugal terms projected onto the corresponding joint \(J_{1j}\) through $S_j$ are
\[
h_j
= S_j^T
\Big(
\big(M^0_{B_j}\dot{S}_j + \dot{M}^0_{B_j} S_j\big)\dot{q}_j
+ \dot{M}^0_{B_j} V^0_1
\Big).
\]

\paragraph{Gravitational terms}
The gravitational wrench acting at the center of mass of body $B_j$ is
\[
W^0_{\mathrm{grav},B_j}
= M^0_{B_j} G^0 ,
\qquad
G^0 = \operatorname{col}(0_5,-g_r),
\]
where $g_r$ is the gravitational acceleration.
The base component $g_b$ and \(B_j\)'s gravitational wrench projected onto the corresponding joint \(J_{1j}\) through $S_j$ are respectively calculated by
\[
g_b = -\sum_{j=1}^{5} W^0_{\mathrm{grav},B_j},
\qquad
g_p = - S_p^T \operatorname{col}(W^0_{\mathrm{grav},B_j}).
\]

\paragraph{Applied external wrenches}
The external wrench applied at the base $B_1$'s center of mass and expressed in the world frame is
\[
W^0_{\mathrm{app},b}
= \sum_{j=2}^{5} W^0_{\mathrm{app},B_j}=\sum_{j=2}^{5}\operatorname{Ad}^T_{(C^0_{f_j})^{-1}}\bar{f}_j 
\]
where \(W_{app,B_j}^0 \in {se}(3)^*\) represents the applied spatial wrench at \(B_j\)'s CoM. The external wrenches acting on the pendulum links $B_2,B_3,B_4$ projected onto the joints are
\[
W^0_{\mathrm{app},p}
= S_p^T \operatorname{col}(W^0_{\mathrm{app},B_j}) .
\]
The total applied wrench vector is therefore
\[
W^0_{\mathrm{app}}
= \big( W^0_{\mathrm{app},b} \;\; W^0_{\mathrm{app},p} \big)^T .
\]
\subsection{Efficient Implementation of the Controller}
  \label{ABASection}
We recall here the modified Articulated-body Inertia Algorithm (ABA) \cite{lieahmed,featherstone2014rigid}, which is utilized to recursively obtain the acceleration $(\ddot{q}_t, \dot{V}^0_1)$ needed to implement the feedback linearizing law~\eqref{lineariningW}~\eqref{Virtual} efficiently, instead of inverting the mass matrix $M$ in~\eqref{statespace}. 

For the system in Fig. \ref{main_vehicle}, a  modified ABA computes $(\ddot{q}_t, \dot{V}^0_1)$ executing the following computations for each $j=2,3,4$
\begin{equation} 
    \begin{split}
        {M}_{j2}^A &= {M}^0_{B_j}  \\
        \psi_{j2}&=\left( S_{j2}^T M^A_{j2} S_{j2} \right)^{-1}\\
        {W}_{j2}^A &=  -\operatorname{ad}^T_{{V}_{j2}^0}{M}^0_{B_j}{V}^0_{j2} - W_{app,B_j}^0 - W_{grav,B_j}^0\\
        {M}_{j1}^A &=   M_{j2}^A - M_{j2}^A S_{j2} \psi_{j2} S_{j2}^T M_{j2}^A  \\
        \psi_{j1}&=\left( S_{j1}^T M^A_{j1} S_{j1} \right)^{-1}\\
        {W}_{j1}^A &=   M_{j2}^A \left( S_{j2} \ddot{\Tilde{q}}_{j2} + \dot{S}_{j2} \dot{q}_{j2} \right) + {W}_{j2}^A
    \end{split}
\end{equation}
where $\ddot{\Tilde{q}}_{ji}$, with $i=1,2$, is given by:
    \begin{equation} 
        \begin{split}
            \ddot{\Tilde{q}}_{ji} &= \psi_{ji} \left( \tau_{f_{ji}} - S_{ji}^T \left( M^A_{ji} \dot{S}_{ji} \dot{q}_{ji}  +{W}_{ji}^A\right) \right)
        \end{split}
    \end{equation}
Afterwards, the articulated inertia $ {M}_{1}^A$ and bias wrench $ {W}_{1}^A$ for the base are computed as
\begin{equation}
    \resizebox{1\columnwidth}{!}{$
\begin{split}
    {M}_{1}^A &=  {M}^0_{B_1} +{M}^0_{B_5} + \sum_{j=2}^{4} \left( M_{j1}^A - M_{j1}^A S_{j1} \psi_{j1} S_{j1}^T M_{j1}^A \right), \\
    {W}_{1}^A &=  - \operatorname{ad}^T_{{V}_1^0}({M}^0_{B_1} + {M}^0_{B_5}){V}^0_{1} - W_{app,B_1}^0 - W_{grav,B_1}^0 - W_{grav,B_5}^0 \\ 
    &\quad + \sum_{j=2}^{4} \big(M_{j1}^A ( S_{j1} \ddot{\Tilde{q}}_{j1} + \dot{S}_{j1} \dot{q}_{j1} ) + {W}_{j1}^A\big).
\end{split}  
$}
\end{equation}
Hence, the following six linear equations are solved for the base acceleration twist $\dot{V}_1^0$ using $LDL^T$ method, since $M_1^A$ is symmetric positive definite
\begin{equation}
0 = M_1^A \dot{V}_1^0 + {W}_1^A.   
\end{equation}
The obtained value of $\dot{V}_1^0$ is then used to retrieve the joint accelerations by evaluating the following expressions:
 \begin{equation}
\begin{split}
    \ddot{q}_{j1} &= -\psi_{j1}  S_{j1}^T M^A_{j1} \dot{V}_{1}^0+\ddot{\Tilde{q}}_{j1}\\
    \dot{V}_{j1}^0 &= \dot{V}_{1}^0 + S_{j1} \ddot{q}_{j1} + \dot{S}_{j1} \dot{q}_{j1}\\
    \ddot{q}_{j2} &= -\psi_{j2}S_{j2}^T M^A_{j2} \dot{V}_{j1}^0 +\ddot{\Tilde{q}}_{j2}.
\end{split}
\end{equation}
\subsection{Lemma 1 and its proof}
\label{LemmaProof}
\begin{lemma}
\label{Lemma1}
Consider a vehicle with a thrust magnitudes vector $f=col(f_i)$ $\in \mathbb{R}_{\geq 0}^l$ with the corresponding propellers configurations $C^0_{i}(C^0_{1},\alpha_i)\in SE(3)$, relative to a right-handed inertial frame whose z-axis direction opposes the axis along which the gravitational force is applied, where $C^0_{1}\in SE(3)$ is the base configuration and $\alpha_i \in \mathbb{S}^2$ is the unit direction vector of the propeller $i$ expressed in the inertial frame. At any equilibrium for the MAV with a given $\bar{C}^0_{1}$, there is no internal force generated at such equilibrium if and only if all propellers, whose thrust magnitudes are $f_i>0$, are pointing in the direction of inertial frame's z-axis, i.e. $C^0_{i}(\bar{C}^0_{1},\alpha_i)\,e_3=e_3$ $\forall i \in \{1, \dots, l\}$ with $e_3=(\,0\, 0\, 1\, 0\,)^T$. 
\end{lemma}
     \subsubsection{Necessity Proof} Assume that at least one propeller, labeled $l$, is not vertical in the inertial frame, hence $C^0_{l}(\bar{C}^0_{1},\alpha_l)\,e_3\neq e_3$. The thrust force vector of propeller $i$, expressed in the inertial frame, is denoted by $T^0_i=f_i\,\alpha_i$. When the MAV is at equilibrium, the resultant thrust force in the inertial frame must satisfy $T^0_{tot}=m_{\rm tot}g_r\,(\,0\, 0\, 1\,)^T$ to compensate for the gravitational force. Thus, this constraint must hold at equilibrium
 \begin{align}
    & \resizebox{0.88\columnwidth}{!}{$\sum_{i=1}^{l-1} f_i\,(0\;0\;1)^T + f_l\,(\alpha_{lx}\;\alpha_{ly}\;\alpha_{lz})^T = m_{\rm tot}g_r\,(0\;0\;1)^T$} \label{constlemma} \\
    & \sqrt{\alpha_{lx}^2 + \alpha_{ly}^2 + \alpha_{lz}^2} = 1 \label{alphaunit}
\end{align}
Expanding this yields 
\begin{equation}
        {f}_l\,\alpha_{lx}=0,\,\,  {f}_l\,\alpha_{ly}=0,\,\, {f}_l\,\alpha_{lz}=m_{\rm tot}g_r- \sum^{l-1}_{i=1}f_i
        \label{lemmaproof1}
\end{equation}
If ${f}_l>0$ then~\eqref{lemmaproof1} gives $(\,\alpha_{lx},\, \alpha_{ly}\,)=(0,0)$. This makes~\eqref{alphaunit} have only two possible solution $\alpha_{lz}=\pm1$. Setting $\alpha_{lz}=1$ in~\eqref{lemmaproof1} yields $\sum^{l-1}_{i=1}f_i<m_{\rm tot}g_r$. This solution for $\alpha_{l}$ is in a contradiction with the assumption as it implies that $C^0_{l}(\bar{C}^0_{1},\alpha_l)\,e_3= e_3$ is true, i.e. propeller $l$ is vertical in the inertial frame, and~\eqref{lemmaproof1} admits a solution of $f_i=\frac{1}{l}m_{\rm tot}g_r$. As a result, $f\in\mathcal{F}_{\rm min}^{\rm eq}$ and no internal force is generated at equilibrium. If instead ${f}_l>0$ and $(\,\alpha_{lx},\, \alpha_{ly},\,\alpha_{lz}\,)=(0,0,-1)$, then from~\eqref{lemmaproof1} we have $\sum^{l-1}_{i=1}f_i>m_{\rm tot}g_r$. While this is not a contradiction, it leads to $f\in\mathcal{F}_{\rm int.for.}^{\rm eq}$. Hence, the assumption $C^0_{l}(\bar{C}^0_{1},\alpha_l)\,e_3\neq e_3$ does not hold when $f\in\mathcal{F}_{\rm min}^{\rm eq}$, proving the necessary condition by contradiction. By induction, the same steps can be easily repeated to cases where two or more propellers assume $C^0_{l}(\bar{C}^0_{1},\alpha_l)\,e_3\neq e_3$.
\subsubsection{Sufficiency Proof} If all propellers point up at equilibrium, i.e. $C^0_{i}(\bar{C}^0_{1},\alpha_i)\,e_3=e_3$ $\forall i \in \{1, \dots, l\}$, then this renders all direction vectors $\alpha_i$'s equal to $(\,0\, 0\, 1\,)^T$, which automatically satisfies~\eqref{constlemma} with $f_i=\frac{1}{l}m_{\rm tot}g_r$. Hence, $f\in\mathcal{F}_{\rm min}^{\rm eq}$.

\subsection{Proof of Proposition 1}
\label{proofProp1}
The proof is constructive and divided into two parts. In the first part, Statement 1 is proven by finding PJD$_2$ joint equilibrium configurations ${\color{black}{q_{pe}}}^{[2]}$ from~\eqref{Case2static1} for any given base pose $C^0_{1e}\in SE(3)$ and showing that the associated equilibrium thrust $f_{e}^{[2]}$ belongs to the minimal  equilibrium thrust $\mathcal{F}_{\rm min}^{\rm eq}$ defined in~\eqref{minDefinition_Fmin} and remains fixed for any such pose $C^0_{1e}$, rendering Statement 2 an immediate corollary as the condition $f_{e}^{[2]}\in\mathcal{F}_{\rm min}^{\rm eq}$ implies that no excess thrust is present at any $C^0_{1e}$, i.e. $f_{e}^{[2]}\notin \mathcal{F}_{\rm int.for.}^{\rm eq}$ (Definition \ref{Def1} \eqref{minDefinition_Fmin}).

In the second part, we start by solving~\eqref{static1} for ${\color{black}{q_{pe}}}^{[1]}$. We then illustrate that, when $\tau_{dj}\neq 0$, there is no joint equilibrium satisfying $C^0_{j}(C^0_{1e},q^{[1]}_{je})e_3= e_3$, hence by Lemma \ref{Lemma1}, no minimal thrust is produced at these euilibria $q^{[1]}_{pe}$, which is the subject of statement 3. Once again, statement 4 follows as a result since $C^0_{j}(C^0_{1e},q^{[1]}_{je})e_3\neq e_3$ necessarily implies $f_{e}^{[1]}\in \mathcal{F}_{\rm int.for.}^{\rm eq}$. However, we observe that $f_{e}^{[1]}$, when $\tau_{dj}\neq 0$, $f_{e}^{[1]}$ can be made arbitrarily close to the set $\mathcal{F}_{\rm min}^{\rm eq}$~\eqref{minDefinition_Fmin}, i.e. ${\color{black}{q_{pe}}}^{[1]}$ is arbitrarily close to meet $C^0_{j}(C^0_{1e},q^{[1]}_{je})e_3= e_3$, by changing the link parameters. Lastly, we give the definitions of the equilibria sets for the state and thrust.

\subsubsection{Statement 1 Proof}
\label{Statement 1 Proof Prop}
Expanding the LHS of~\eqref{Case2static1} or~\eqref{static1} for any $B_j$ $\forall j=2,3,4$ yields the pair of equations for $B_j$
\begin{equation}
    {\color{black}{g_{p_j}}}=-S^T_{j} W_{grav,B_j}^0 =(\,\,-S^T_{j1} \,\,-S^T_{j2}\,\,)^T\,
    M_{B_j}^0 G^0
    \label{g_tj}
\end{equation}
where the instantaneous spatial joint screws are given by $S_{j1}(C^0_1)=\operatorname{Ad}_{C^0_1}Y_{j1}$ and $S_{j2}(C^0_1,q_{j1})=\operatorname{Ad}_{C^0_{j1}}\operatorname{Ad}^{-1}_{A^0_{j1}}Y_{j2}$, with $A^0_{j1}$ being the configuration of the intermediate link $\bar{B}_{j}$ frame, placed at the center of rotation of $q_{j1}$, connecting $B_1$ and $B_{j}$ when the vehicle is in the home configuration shown in Fig. \ref{main_vehicle}, and $C^0_{j1}$ being its configuration in a generic vehicle pose. $Y_{ji} \in \mathbb{R}^{6}$ is the constant spatial screw of $q_{ji}$ in home configuration. For details on screw theory, consult~\cite{Murray2017Mathematical}.

Now, consider the LHS of first equation in~\eqref{g_tj}, which is written as
\begin{equation}
\begin{split}
     g_{p_{j1}}&=-S^T_{j1}M_{B_j}^0 (\,\,0_{1\times5} \,\,-{g_r}\,\,)^T\\
     &= Y^T_{j1}\operatorname{Ad}^T_{C^0_1}\operatorname{Ad}^{-T}_{C^0_{j}}I^{j}_{B_j}\operatorname{Ad}^{-1}_{C^0_{j}}\bar{e}^6_6 g_r
\end{split}
\label{EqSj1}
\end{equation}
where $I^{j}_{B_j}$ is the diagonal $B_j$'s CoM-referred inertia tensor. $\bar{e}^k_i\in\mathbb{R}^k$ is a zero vector with $1$ in the $i$-th component. Since the adjoint matrix $\operatorname{Ad}^{-T}_{C^0_{j}}$ has the properties
\begin{equation}
\begin{split}
        \operatorname{Ad}^{-T}_{C^0_{j}}&=\operatorname{Ad}^{T}_{C^j_{0}}=(\operatorname{Ad}_{C^j_{1}C^1_{0}})^{T}=(\operatorname{Ad}_{C^j_{1}}\operatorname{Ad}_{C^1_{0}})^{T}\\
        &=\operatorname{Ad}^{T}_{C^1_{0}}\operatorname{Ad}^{T}_{C^j_{1}}=\operatorname{Ad}^{-T}_{C^0_{1}}\operatorname{Ad}^{-T}_{C^1_{j}}
        \label{adjprop}
\end{split}
\end{equation}
substituting $\operatorname{Ad}^{-T}_{C^0_{j}}=\operatorname{Ad}^{-T}_{C^0_{1}}\operatorname{Ad}^{-T}_{C^1_{j}}$ in~\eqref{EqSj1} gives
\begin{equation}
        g_{p_{j1}}=Y^T_{j1}\operatorname{Ad}^{-T}_{C^1_{j}}I^{j}_{B_j}\operatorname{Ad}^{-1}_{C^0_{j}}\bar{e}^6_6 g_r=Y^T_{j1}\operatorname{Ad}^{-T}_{C^1_{j}}I^{j}_{B_j}\bar{r}_j g_r
        \label{Sj1_r}
\end{equation}
where $\bar{r}_j$ denotes
\begin{equation}
    \bar{r}_j=\operatorname{Ad}^{-1}_{C^0_{j}}\bar{e}^6_6 \label{r_exp}
\end{equation}

Observe that substituting $\bar{r}_j=\bar{e}^6_6$ in~\eqref{Sj1_r} reduces it to
\begin{equation}
\begin{split}
    g_{p_{j1}}&=Y^T_{j1}\operatorname{Ad}^{-T}_{C^1_{j}}I^{j}_{B_j}\bar{e}^6_6 g_r\\
    &=Y^T_{j1}\operatorname{Ad}^{-T}_{C^1_{j}}\bar{e}^6_6 m_p g_r
\end{split}
\label{Sj1_reduced}
\end{equation}
where $g_{p_{j1}}$ now depends only on $q_j$. Since the constant screw is $Y^T_{j1}=(\,\,(\bar{e}^3_1)^T\,\,\,\sigma_j \frac{a}{2}(\bar{e}^3_3)^T\,\,)$, where $\sigma_j=1$ for $j=2,5$ and $\sigma_j=-1$ otherwise, \eqref{Sj1_reduced} can be manipulated further as 
\begin{equation}
\resizebox{0.89\columnwidth}{!}{$
\begin{split}
    g_{p_{j1}}&=(\,(\bar{e}^3_1)^T \;\; \sigma_j \tfrac{a}{2}(\bar{e}^3_3)^T\,)
    \begin{pmatrix}
        R^1_{j} & -(P^1_{j})_{\times}^T R^1_{j} \\ 0_{3\times3} & R^1_{j}
    \end{pmatrix}
    \bar{e}^6_6\, m_p g_r\\
    &=(\,(\bar{e}^3_1)^T \;\; \sigma_j \tfrac{a}{2}(\bar{e}^3_3)^T\,)
    \begin{pmatrix}
        -(P^1_{j})_{\times}^T R^1_{j} \bar{e}^3_3 \\ R^1_{j} \bar{e}^3_3
    \end{pmatrix}
    m_p g_r\\
    &=(\sigma_j \tfrac{a}{2}(\bar{e}^3_3)^T - (\bar{e}^3_1)^T (P^1_{j})_{\times}^T) R^1_{j} \bar{e}^3_3\, m_p g_r\\
    &=0\;\equiv\; \text{RHS of first equation in \eqref{Case2static1}.}
\end{split}
$}
\end{equation}
where $C^1_{j}=\left(\begin{smallmatrix}
        R^1_{j} & P^1_{j} \\ 0_{1\times3}& 1
    \end{smallmatrix}\right)$. Hence setting $\bar{r}_j=\bar{e}^6_6$ satisfies the first equation in \eqref{Case2static1}.

Next, let us consider the second equation in \eqref{Case2static1}. Simplifying using the properties in~\eqref{adjprop}, and substituting the expression of the constant screw $Y_{j2}$, configuration $A^0_{j1}$ and $\bar{r}_j$~\eqref{r_exp} yield
\begin{equation}
\begin{split}
     g_{p_{j2}}&= Y_{j2}^T\operatorname{Ad}^{-T}_{A^0_{j1}}\operatorname{Ad}^T_{C^0_{j1}}\operatorname{Ad}^{-T}_{C^0_{j}}I^{j}_{B_j}\operatorname{Ad}^{-1}_{C^0_{j}}\bar{e}^6_6 g_r\\
     &=Y_{j2}^T\operatorname{Ad}^{-T}_{A^0_{j1}}\operatorname{Ad}^{-T}_{C^{j1}_{j}}I^{j}_{B_j}\bar{r}_j g_r\\
     &=(\,\,(\bar{e}^3_2)^T\,\,0_{1\times3}\,\,)\operatorname{Ad}^{-T}_{C^{j1}_{j}}I^{j}_{B_j}\bar{r}_j g_r
\end{split}
\label{EqSj2}
\end{equation}
Setting $\bar{r}_j=\bar{e}^6_6$ gives
\begin{align}
     g_{p_{j2}}&=(\,\,(\bar{e}^3_2)^T\,\,0_{1\times3}\,\,)\operatorname{Ad}^{-T}_{C^{j1}_{j}}\bar{e}^6_6 m_p g_r\\
     &=(\,\,(\bar{e}^3_2)^T\,\,0_{1\times3}\,\,)\left(\begin{smallmatrix}
         -(P^{j1}_{j})_{\times}^TR^{j1}_{j}\bar{e}^3_3 \\  R^{j1}_{j}\bar{e}^3_3
    \end{smallmatrix}\right)m_p g_r \nonumber\\
    &=-(\bar{e}^3_2)^T(P^{j1}_{j})_{\times}^TR^{j1}_{j}\bar{e}^3_3 \,m_p g_r \nonumber\\
    &=\,(\,\,\,-\cos(q_{j2e})\,\,\,0\,\,\,\sin(q_{j2e}) \,\,\,)\left(\begin{smallmatrix}
         \sin(q_{j2e}) \\  0\\ \cos(q_{j2e})
    \end{smallmatrix}\right)c \,m_p g_r \nonumber\\
    &=0 \text{ $\equiv$ RHS of second equation in \eqref{Case2static1}.} \nonumber
    \label{EqSj2_2}
\end{align}
We conclude that $\bar{r}_j=\bar{e}^6_6$ satisfies \eqref{Case2static1}. Observe that this condition implies that all propellers are aligned vertically in the inertial frame at equilibria for any $C^0_{1e}\in SE(3)$. To see this, expand \eqref{r_exp} as
\begin{align}
    \bar{r}_j&=\bar{e}^6_6
    =\operatorname{Ad}^{-1}_{C^0_{j}}\bar{e}^6_6\\
    \left(\begin{smallmatrix}
        0_{3\times 1} \\ \bar{e}^3_3    \end{smallmatrix}\right)&=\left(\begin{smallmatrix}
        R^j_{0} &  0_{3\times3}\\ -R^j_{0}(P^0_{j})_{\times} & R^j_{0}
    \end{smallmatrix}\right)\left(\begin{smallmatrix}
        0_{3\times 1} \\ \bar{e}^3_3
    \end{smallmatrix}\right)=\left(\begin{smallmatrix}
        0_{3\times 1} \\ R^j_{0}\bar{e}^3_3
    \end{smallmatrix}\right) \nonumber
\end{align}
Hence $\bar{e}^3_3=R^j_{0}(R^0_{1e},q^{[2]}_{je})\bar{e}^3_3 \,\,\forall j=2,3,4,5$, or alternatively $C^0_{j}e_3= e_3$ using Lemma \ref{Lemma1} notation. This also gives the expression of $q^{[2]}_{je}$ by solving $\bar{r}_j=\bar{e}^6_6$ for $q^{[2]}_{je}$ as
\begin{equation}
\begin{split}
        R^{1e}_{0}\bar{e}^3_3= R^{1}_{j}(q^{[2]}_{je})\bar{e}^3_3=R_x(q^{[2]}_{j1e})R_y(q^{[2]}_{j2e})\bar{e}^3_3
\end{split} \label{R0prop}
\end{equation}
Thus we have the solution $q^{[2]}_{je}$ as
\begin{equation}
\begin{split}
    l_{s1}&=\frac{-r_{32}}{\sqrt{1-r_{31}^2}},\,l_{c1}=\sqrt{1-l_{s1}^2}\\
    l_{s2}&=r_{31},\,l_{c2}=\sqrt{1-l_{s2}^2}\\
    q^{[2]}_{jie}&=\operatorname{atan}(l_{si},l_{ci})\,\, \forall i=1,2, \,j=2,3,4,5
\end{split} \label{PJD2Equi_Joint}
\end{equation}
where $r_{mn}$ is the element $(m,n)$ in any desired base orientation at PJD$_2$ equilibrium $R^{0}_{1e}$. Notice that $q^{[2]}_{je}$ is equal for any attached link $B_j$. Moreover, it is always possible to obtain this solution $q^{[2]}_{je}$ \eqref{PJD2Equi_Joint} for every $R^{0}_{1} \in SO(3)$ since aligning the unit vector $R^{1}_{j}(q^{[2]}_{je})\bar{e}^3_3=:z^1_j$, where $z^1_j\in \mathbb{S}^2$ is the z-axis of $\mathcal{F}_{p_j}$ expressed in $\mathcal{F}_{b}$, with any desired unit vector $z^1_0\in \mathbb{S}^2$ in $\mathcal{F}_{b}$, where $R^{1}_{0}\bar{e}^3_3=:z^1_0$, requires only 2 variables, i.e. $q^{[2]}_{j1e},q^{[2]}_{j2e}$, as $\operatorname{dim}(\mathbb{S}^2)=2$ when the manifold $\mathbb{S}^2$ is embedded in the 3-dimensional Euclidean vector space.

Now, we show that the associated thrust $f^{[2]}_{e} \in \mathcal{F}_{\rm min}^{\rm eq}$ and fixed for any $C^{0}_{1e}$. The dynamics of base at any static equilibrium reduces to
\begin{equation}
\begin{split}
    g_b(C^0_{1e},{\color{black}{q_{pe}}})=&-W^0_{app,b}(f_e,{\color{black}{q_{pe}}},{\color{black}{q_{ae}}}) 
\end{split}
\label{static2}
\end{equation} which reads explicitly as 
\begin{equation}
\sum_{j=2}^{5}\operatorname{Ad}^T_{(C^0_{f_j})^{-1}}\Bar{f}_j f_{je}= \sum_{j=1}^{5}\operatorname{Ad}^{-T}_{(C^0_{j})}I^{j}_{B_j}\operatorname{Ad}^{-1}_{(C^0_{j})}\bar{e}^6_6 g_r
     \label{static3}
\end{equation}
Substituting $\bar{r}_j=(\,\,0_{3\times1}\,\, r_{g_j}\,\,)$ from \eqref{r_exp} yields
\begin{equation}
\resizebox{0.89\columnwidth}{!}{$
\begin{split}
        \sum_{j=2}^{5} 
        \begin{pmatrix}
        (I k_j - (P^1_{j})_{\times}^T) R^1_{j} \bar{e}^3_3 \\ 
        R^1_{j} \bar{e}^3_3
        \end{pmatrix} 
        f_{je}
        &=
        g_r\, m_b 
        \begin{pmatrix}
        0_{3\times 1} \\ 
        R^1_{0} \bar{e}^3_3
        \end{pmatrix} \\
        &\quad
        + 
        g_r \sum_{j=2}^{5} m_j
        \begin{pmatrix}
        -(P^1_{j})_{\times}^T R^1_{j} r_{g_j} \\
        R^1_{j} r_{g_j}
        \end{pmatrix}
\end{split}
$}
\label{eq_base_2}
\end{equation}
where $I$ is the identity matrix. Using the property $R^1_{j}(R^j_{1}P^1_{j})_{\times}=-(P^1_{j})_{\times}^TR^1_{j}$ and $R^1_{j}r_{g_j}=R^1_{0}\bar{e}^3_3$, we get 
\begin{equation}
\resizebox{0.89\columnwidth}{!}{$
\begin{split}
  &\sum_{j=2}^{5}R^1_{j}( Ik_j+(R^j_{1}P^1_{j})_{\times})\bar{e}^3_3 f_{je}=g_r\Big(\sum_{j=2}^{5}R^1_{j}
         (R^j_{1}P^1_{j})_{\times}r_{g_j}m_j\Big)\\
        &\sum_{j=2}^{5} R^1_{j}\bar{e}^3_3f_{je}=g_r\sum_{j=2}^{5}R^1_{j}r_{g_j}\Big(\frac{1}{4}m_b+m_j\Big)
\end{split}$}
\label{GeneralThrust}
\end{equation}
Since in PJD$_2$, $q^{[2]}_{je}\, \forall j=2,3,4,5$ are equal, $R^1_{j}$ can be taken as a common factor, hence pre-multiplying by $R^j_{1}$ gives
\begin{equation}
\resizebox{0.89\columnwidth}{!}{$
\begin{split}
  & \sum_{j=2}^{5}k_jf^{[2]}_{je}\bar{e}^3_3 +\sum_{j=2}^{5}(R^j_{1}P^1_{j})_{\times}f^{[2]}_{je}\bar{e}^3_3 =g_r\Big(\sum_{j=2}^{5}
         (R^j_{1}P^1_{j})_{\times}m_j\Big)\bar{e}^3_3\\
        &\bar{e}^3_3\sum_{j=2}^{5} f^{[2]}_{je}=\bar{e}^3_3\sum_{j=2}^{5}g_r\Big(\frac{1}{4}m_b+m_j\Big)
\end{split}$}
\end{equation}
After $r_{g_j}=\bar{e}^3_3$ is substituted. The second equation is satisfied with $f^{[2]}_{je}={g_r(\frac{1}{4}m_b+m_j)}$, hence $f^{[2]}_{e} \in \mathcal{F}_{\rm min}^{\rm eq}$. Substituting this $f^{[2]}_{je}$ make the first equation reduce to 
\begin{equation}
    \sum_{j=2}^{5} k_j{g_r(\frac{1}{4}m_b+m_j)}\bar{e}^3_3+\sum_{j=2}^{5}(R^j_{1}P^1_{j})_{\times}\bar{e}^3_3=0
\end{equation} thus $\sum_{j=2}^{5} k_j{(\frac{1}{4}m_b+m_j)}=0$ since $\sum_{j=2}^{5}(R^j_{1}P^1_{j})_{\times}\bar{e}^3_3=0$ is always true at any $q^{[2]}_{je}$ \eqref{PJD2Equi_Joint}. We conclude that the associated thrust at $q^{[2]}_{je}$ is, when $m_j=m_p\, \forall j=2,3,4,5$
\begin{equation}
    f^{[2]}_{je}={g_r(\frac{1}{4}m_b+m_p)},\text{ with $k_i$ chosen by}\sum_{j=2}^{5}k_i=0
    \label{PJD2Thrust}
\end{equation}
\subsubsection{Statement 2 Proof} Since $f^{[2]}_{e} \in \mathcal{F}_{\rm min}^{\rm eq}$ is proven in Statement 1, it follows from Definition \ref{Def1} that no internal forces are generated $f^{[2]}_{e} \notin \mathcal{F}_{\rm int.for.}^{\rm eq}$ at the corresponding equilibrium configurations associated with this $f^{[2]}_{e}$, i.e. $q^{[2]}_{te}$ in \eqref{PJD2Equi_Joint} and any desired $C^0_{1e} \in SE(3)$. 
\subsubsection{Statement 3 Proof}
In Sec.~\ref{Statement 1 Proof Prop}, it is found that when all propellers point up in the inertial frame, i.e. $\bar{r}_j=\bar{e}^6_6 \text{ or } C^0_{j}e_3= e_3\,\,\forall j=2,3,4,5$, the LHS of~\eqref{static1} becomes ${\color{black}{g_{p_j}}}(C^0_{1e},{\color{black}{q_{pe}}})=0$. Since the RHS of of it for any $j$ is equal to $\left(\begin{smallmatrix}
         \tau_{dj} \\  0
    \end{smallmatrix}\right)$, which is not zero when $\tau_{dj}\neq0$, $C^0_{j}(C^0_{1e},q^{[1]}_{je})e_3\neq e_3$ must hold true. Thus, from Lemma \ref{Lemma1}, no minimal thrust is produced at any PJD$_1$ equilibrium $f^{[1]}_{e}\notin\mathcal{F}_{\rm min}^{\rm eq} \,\forall\tau_{dj}\neq0$.

For completeness, we proceed to find conditions on $q^{[1]}_{pe}$, similar to $\bar{r}_j=\bar{e}^6_6$ obtained for PJD$_2$, that must be met at any base equilibrium $C^0_{1e}$. We start from   \eqref{EqSj2} and substitute the values of the constant screws, constant transformations and $\bar{r}_j=(\,\,0_{3\times1}\,\, r_{g_j}\,\,)$, which yields
    \begin{equation}
       -(\bar{e}^3_2)^T(P^{j1}_{j})_{\times}^TR^{j1}_{j}r_{g_j}   
    m_j g_r= 0
\end{equation}
which is satisfied when $r_{g_j}=(\,\,0\,\, r_{g_j,1}\,\,r_{g_j,2}\,\,),\,r_{g_j,2}=\sqrt{1-r_{g_j,1}^2}$. in PJD$_1$, \eqref{Sj1_r} has the RHS equal to $S^T_{j1}W_{app,B_j}^0$. This can be expressed as 
\begin{equation}
\begin{split}
        (\sigma_j \frac{a}{2}(\bar{e}^3_3)^T-(\bar{e}^3_1)^T(P^1_{j})_{\times}^T)&R^1_{j}\big(r_{g_j}m_j g_r-\bar{e}^3_3 f^{[1]}_{je}\big)\\&+(\bar{e}^3_1)^TR^1_{j}\bar{e}^3_3k_jf^{[1]}_{je}=0
\end{split}
\end{equation}
since
\begin{equation}
    Y^T_{j1}\operatorname{Ad}^{-T}_{C^1_{j}}=\left(\begin{smallmatrix} \cos(q_{j2e})& 0& \sin(q{j2e})& 0& c \cos(q{j2e})& 0\end{smallmatrix}\right)
\end{equation}
hence $q^{[1]}_{j2e}$ is constraint by
\begin{equation}
\cos(q^{[1]}_{j2e})=\pm\sqrt{\frac{k_j^2(f^{[1]}_{je})^2}{c^2m_j^2g_r^2 r_{g_j,1}^2+k_j^2(f^{[1]}_{je})^2}}
\end{equation}
Lastly, from \eqref{GeneralThrust}, we have these conditions on $R^1_{j}(q^{[1]}_{je}),\,f^{[1]}_{je}$ 
\begin{equation}
\begin{split}
&\sum_{j=2}^{5}R^1_{j} k_j\bar{e}^3_3f^{[1]}_{je}=0,\,\,R^1_{j}r_{g_j}=R^1_{0}\bar{e}^3_3\\
  &\sum_{j=2}^{5} (P^1_{j})_{\times}(R^1_{j}\bar{e}^3_3f^{[1]}_{je}-R^1_{j}r_{g_j}g_rm_j)=0\\
        &\sum_{j=2}^{5} R^1_{j}\bar{e}^3_3f^{[1]}_{je}-\sum_{j=2}^{5}R^1_{j}r_{g_j}g_r\Big(\frac{1}{4}m_b+m_j\Big)=0
\end{split}
\label{Case2Equilb_new}
\end{equation}
\subsubsection{Statement 4 Proof}
\label{EquSets}
By definition \ref{Def1}, and Statement 3, $f^{[1]}_{e}\in\mathcal{F}_{\rm int.for.}^{\rm eq}$ immediately follows as a corollary.  The amount of internal force can be made arbitrarily small if solutions of \eqref{Case2static1} and \eqref{static1} are made arbitrarily close, which is realized by making the term $\frac{k_j}{cm_j}=0$, i.e. reducing the drag-to-thrust coefficient or increasing the pendulum length or mass.

Finally, we conclude the proposition proof by defining the equilibria state sets as
\begin{equation}
\begin{split}
       \mathbb{D}^{[i]}_{x_e}=&\Big\{x(t)\in \mathbb{X},\,\{t,\Bar{t}\}\in \mathbb{R}_{\geq0} 
   \,\Bigr\rvert \,\forall C_{1e}^0 \in SE(3), \,\forall t\geq\Bar{t}, \\& \quad x (t):=\left(\,
      C_1^0(t),\,
      {\color{black}{q_{p}}}(t),\, {\color{black}{V_1^0,\dot{q}_{p}}}, \,{\color{black}{q_a}}(t)\,\right)^T\\&= \left(\,C_{1e}^0(\Bar{t}),
      {\color{black}{q_{pe}}}^{[i]}(\Bar{t}),\, 0_{12},\,{\color{black}{q_{ae}}^{[i]}}(\Bar{t})\,\right),
\\& {\color{black}{q_{pe}}}^{[i]} \text{ from }\eqref{Case2Equilb_new} \text{ or }\eqref{PJD2Equi_Joint},\text{ for $i=1,2$, respectively}\Big\}
\end{split} \label{stateSet}
\end{equation}
while the associated thrust equilibrium sets are given by
\begin{equation}
\begin{split}
       \mathbb{D}^{[i]}_{f_e}=&\Big\{f(t)\in \mathbb{R}^4,\,\{t,\Bar{t}\}\in \mathbb{R}_{\geq0} 
   \,\Bigr\rvert \,\forall t\geq\Bar{t}, f(t) = f_{e}^{[i]}(\Bar{t}),
\\& f^{[i]}_e \text{ from }\eqref{Case2Equilb_new} \text{ or }~\eqref{PJD2Thrust},\text{ for $i=1,2$, respectively}\Big\} 
\end{split} \label{inputSet}
\end{equation}
\subsection{Proof of Theorem 1}
\label{proofTheorem1}
The proof relies on the relative degree notion for MIMO affine nonlinear systems~\cite{isidori2013nonlinear} and the dynamic extension algorithm for invertible systems~\cite{DEA} to prove Statement 1. Afterwards, we show that Statement 2 is true by a standard Lyapunov argument applied to the zero dynamics in the normal form. 
\subsubsection{Statement 1 Proof}
To compute the relative degree of $\Sigma_2$ w.r.t. ${\color{black}{V_1^0}}$, we take time derivatives of the output ${\color{black}{V_1^0}}$
\begin{equation}
\begin{split}
       \dot{h}_y=  \dot{V}^0_1  
\end{split}
\end{equation}
From~\eqref{statespace} $\dot{V}^0_1$ is expressed as
\begin{equation}
    M_{bb}\dot{V}^0_{1}=- M_{bp}\ddot{q}_{p}-h_b-g_b+W^0_{app,b} \label{MbbV1dot}
\end{equation}
Following the approach of~\cite{Spong} we can express $\ddot{q}_{p}$ in terms of $\dot{V}^0_{1}$, ${\color{black}{W^0_{app,p}}}$ and ${\tau}_{f}$, hence
\begin{equation}
\begin{split}
         \ddot{q}_{p}=&{\color{black}{L_{1}}}-M_{pp}^{-1} M_{bp}^T \dot{V}^0_{1}+ M_{pp}^{-1} {\color{black}{W_{app,p}^0}}
\end{split}   
\end{equation}
where ${\color{black}{L_1}}$ is defined as
\begin{equation}
    {\color{black}{L_{1}}}:= M_{pp}^{-1}(\tau_{f}-h_t -{\color{black}{g_p}}) \label{TermsCt1}
\end{equation} 
Substituting in~\eqref{MbbV1dot} gives
\begin{equation}
\begin{split}
        (M_{bb}- M_{bp}M_{pp}^{-1} M_{bp}^T ) \dot{V}^0_{1}=&- M_{bp}  {\color{black}{L_{1}}} -h_b-g_b\\
        &+W^0_{app,b}- M_{bp} M_{pp}^{-1} {\color{black}{W_{app,p}^0}} 
\end{split}
 \label{MbbV1dot_2}
\end{equation}
which can be compactly rewritten as 
\begin{equation}
    \dot{h}_y=\dot{V}^0_{1}= E_1+M_b^{-1}D_1 u 
    \label{accZD}
\end{equation}
where $M_b$, $D_1$ and $E_1$ are defined as, with $j=2,3,4$
\begin{equation}
\begin{split}
M_b=&(M_{bb}- M_{bp}M_{pp}^{-1} M_{bp}^T )\\
{\color{black}{N_j}}=&\big(I_{6}- M^0_{B_j}{S}_{j}(S_{j}^T M^0_{B_j}{S}_{j})^{-1}S_j^T\big)\\
\Bar{D}_1=&\left(\begin{array}{ccc}
          {\color{black}{N_j}}\operatorname{Ad}^T_{(C^0_{f_j})^{-1}}\Bar{f}_j & \cdots &\operatorname{Ad}^T_{(C^0_{f_5})^{-1}}\Bar{f}_5 
     \end{array}\right)\\
    D_1=&\left(\begin{array}{cccc}\Bar{D}_1& 0_{6\times2} 
     \end{array}\right)\\
     E_1=&M_b^{-1}(- M_{bp}  {\color{black}{L_{1}}} -h_b-g_b)
\end{split} \label{TermsMb}
\end{equation}
Note that $(S_{j}^T M^0_{B_j}{S}_{j})\in R^{2\times2}$ is a symmetric positive definite matrix.

Since the decoupling matrix $M_b^{-1}D_1$ is singular due to $D_1$ being singular of rank 4  $\forall{x}_e\in \mathbb{D}^{[i]}_{x_e}$, system~\eqref{x2} does not have a vector relative degree in any open neighbourhood of the equilibrium ${x}_e\in \mathbb{D}^{[i]}_{x_e}$. Thus, the system is not I/O feedback linearizable by any static state feedback as the relative degree is invariant under this kind of feedback~~\cite{isidori2013nonlinear}.

However, extending the system dynamics $\Sigma_2$ with an input precompensator, by applying the version of the dynamic extension algorithm presented in~~\cite{DEA} for invertible nonlinear systems, results in a system for which the vector relative degree is well-defined. Assume the precompensator takes this pure integrator form~\eqref{precompensator}.
Let $\bar{x}\in \bar{\mathbb{X}}$ be the extended state and $w\in \mathbb{R}^6$ the new input. The extended system $\Sigma_{\bar{2}}$ can be written as $\Sigma_{\bar{2}}:\dot{\bar{x}}=\bar{F}_2+\bar{G}_2w$ with $\bar{F}:= F_2+G_2 \bar{u}$, where $\bar{u}=(z,0,0)^T$ and $\bar{G}_2$ is given by
\begin{equation}
    \Bar{G}_2=\left(\begin{array}{c}
          0_{18\times6}  \\
          0_{2\times4} \quad I_{2\times2}\\
          I_{4\times4} \quad 0_{2\times2}
     \end{array}\right)
\end{equation}

Now, we can proceed with computing the vector relative degree of $\Sigma_{\Bar{2}}$ w.r.t ${\color{black}{V_1^0}}$ by differentiating $\dot{h}_y$ w.r.t time 
\begin{equation}
    M_{bb}\ddot{V}^0_{1}=- \dot{M}_{bp}\ddot{q}_{p}-{M}_{bp}\dddot{q}_{p}+\dot{W}^0_{app,b}-\dot{M}_{bb}\dot{V}^0_{1}-\dot{h}_b-\dot{g}_b \nonumber
\end{equation}
by substituting this expression for $\dddot{q}_{p}$
\begin{equation}
    \dddot{q}_{p}={\color{black}{L_2}}- M_{pp}^{-1}M_{bp}^T \ddot{V}^0_{1}+ M_{pp}^{-1}{\color{black}{\dot{W}_{app,p}^0}}
\end{equation}
where ${\color{black}{L_2}}$ is defined as
\begin{equation}
    {\color{black}{L_2}}:=M_{pp}^{-1}\big((\dot{\tau}_{f}-{\color{black}{\dot{h}_p}}-{\color{black}{\dot{g}_p}})-( \dot{M}_{bp}^T \dot{V}^0_{1}+\dot{M}_{pp}\ddot{q}_{p})\big)
    \label{TermsCt2}
\end{equation}
and after some algebraic manipulation it yields
\begin{equation}
\begin{split}
      M_b\ddot{V}^0_{1}=&- \dot{M}_{bp}({\color{black}{L_{1}}}+ M_{pp}^{-1} {\color{black}{W_{app,p}^0}})-{M}_{bp}{\color{black}{L_2}}\\&+(\dot{M}_{bp}M_{pp}^{-1} M_{bp}^T -\dot{M}_{bb})\dot{V}^0_{1}-\dot{h}_b-\dot{g}_b \\&-{M}_{bp} M_{pp}^{-1}{\color{black}{\dot{W}_{app,p}^0}}+\dot{W}^0_{app,b}
\end{split} 
\label{AccDec}
\end{equation}
where the $\dot{h}_b, \dot{g}_b,{\color{black}{\dot{h}_p}},{\color{black}{\dot{g}_p}}, \ddot{M},\ddot{S}_{j}, \dot{W}_{app,B_j}^0$ are given by
\begin{equation}
    \resizebox{0.90\columnwidth}{!}{$
    \begin{split}
        \dot{h}_b=&\sum^{4}_{j=2}\big( (M^0_{B_j}\ddot{S}_{j}+\ddot{M}^0_{B_j}{S}_{j}+2\dot{M}^0_{B_j}\dot{S}_{j})\dot{q}_{j}\\&+(M^0_{B_j}\dot{S}_{j}+\dot{M}^0_{B_j}{S}_{j})\ddot{q}_{j}\big)+\ddot{M}_{bb}{V}^0_{1}+\dot{M}_{bb}\dot{V}^0_{1}\\
        {\color{black}{\dot{h}_p}}=&\operatorname{col}\big(\dot{S}_{j}^T((M^0_{B_j}\dot{S}_{j}+\dot{M}^0_{B_j}S_{j})\dot{q}_{j}+\dot{M}^0_{B_j}{V}^0_{1})\\&\quad\quad+S_{j}^T((M^0_{B_j}\ddot{S}_{j}+\ddot{M}^0_{B_j}S_{j}+2\dot{M}^0_{B_j}\dot{S}_{j})\dot{q}_{j}\\&\quad\quad+(M^0_{B_j}\dot{S}_{j}+\dot{M}^0_{B_j}S_{j})\ddot{q}_{j}+\ddot{M}^0_{B_j}{V}^0_{1}+\dot{M}^0_{B_j}\dot{V}^0_{1})\big)\\
        \dot{g}_b&=-(\dot{M}_{bb}+\dot{M}^0_{B_5})G^0\\
        {\color{black}{\dot{g}_p}}&= -{\color{black}{\dot{S}_p^T}} \operatorname{col}({M}^0_{B_j}G^0)-S_p^T \operatorname{col}(\dot{M}^0_{B_j}G^0) \\
        {\color{black}{\ddot{M}^0_{B_j}}}&={\color{black}{{M}^0_{B_j}}} \operatorname{ad}^2_{{V}^0_k}+(\operatorname{ad}^2_{{V}^0_k})^T{\color{black}{{M}^0_{B_j}}}+2\operatorname{ad}^T_{{V}^0_k}{\color{black}{{M}^0_{B_j}}} \operatorname{ad}_{V^0_k}\\&-{\color{black}{{M}^0_{B_j}}} \operatorname{ad}_{\dot{V}^0_k}-\operatorname{ad}^T_{\dot{V}^0_k}{\color{black}{{M}^0_{B_j}}},\text{ if } k=5 \text{ then } {V}^0_k={V}^0_1,\, \dot{V}^0_k=\dot{V}^0_1\\
        \ddot{S}_{k}&= (\operatorname{ad}_{\dot{V}^0_k} +\operatorname{ad}^2_{V^0_k}) {S}_{k},\,k=2,3,4\\
        \dot{W}_{app,B_k}^0&= -\operatorname{ad}^T_{V_k^0}{W}_{app,B_k}^0+\operatorname{Ad}^T_{(C^0_{f_5})^{-1}}\dot{W}_{app,B_k}^b,\,k=2,3,4,5
    \end{split}$} \label{TermsDervi}
\end{equation}
where $j=2,3,4$. Note that ${M}^0_{B_5}$ depends only on $C^0_1$, i.e. ${M}^0_{B_5}(C^0_1)$, due to the assumptions on link $B_5$. Hence, the input ${\color{black}{\dot{q}_a}}$ does not appear in $\dot{M}^0_{B_5}(C^0_1)$. Observe that~\eqref{AccDec} can be compactly put in the form, 
\begin{equation}
    \ddot{h}_y=\ddot{V}^0_{1}= E_2+M_b^{-1}D_2 w
\end{equation}
where $E_2$ and $D_2$ are calculated from, with $j=2,3,4$
\begin{equation}
 \resizebox{0.90\columnwidth}{!}{$
 \begin{aligned}
        D_2&=\left(\begin{array}{cccc} {\color{black}{N_j}}\operatorname{Ad}^T_{(C^0_{f_j})^{-1}}\Bar{f}_j & \cdots &\operatorname{Ad}^T_{(C^0_{f_5})^{-1}}\Bar{f}_5& \Bar{D}_{2} 
     \end{array}\right) \\ 
     \Bar{D}_{2} &= \left(\begin{array}{cc}-\operatorname{ad}^T_{S_{51}}\operatorname{Ad}^T_{(C^0_{f_5})^{-1}}\Bar{f}_5 z_5 & -\operatorname{ad}^T_{S_{52}}\operatorname{Ad}^T_{(C^0_{f_5})^{-1}}\Bar{f}_5 z_5 \end{array}\right)\\
     E_2&= M_b^{-1}\big(- \dot{M}_{bp}({\color{black}{L_{1}}}+ M_{pp}^{-1} {\color{black}{W_{app,p}^0}})-{M}_{bp}{\color{black}{L_2}}\\&\quad\quad\quad\quad+(\dot{M}_{bp}M_{pp}^{-1} M_{bp}^T -\dot{M}_{bb})\dot{V}^0_{1}-\dot{h}_b-\dot{g}_b+\Bar{W}\big)\\
     \Bar{W}&= -\sum^{4}_{j=2}B_j\operatorname{ad}^T_{V_{j2}^0}W_{app,B_j}^0 - \operatorname{ad}^T_{V_{1}^0}W_{app,B_5}^0
\end{aligned}$} \label{TermsW}
\end{equation}
where ${\color{black}{L_{1}}}$, ${\color{black}{L_2}}$, $M_b$ and ${\color{black}{N_j}}$ are given in~\eqref{TermsCt1},~\eqref{TermsCt2} and~\eqref{TermsMb}.

Because of the decoupling matrix $M_b^{-1}D_2$ of the extended system $\Sigma_{\Bar{2}}$ being full rank equal to 6 in the open neighbourhood  $\mathbb{U}_{\bar{x}_e}\subseteq \bar{\mathbb{X}}$  of equilibrium $\bar{x}_e\in \mathbb{D}^{[i]}_{x_e} \cup \mathbb{D}^{[i]}_{f_e}$, system $\Sigma_{\Bar{2}}$ has a vector relative degree $(2,2,2,2,2,2)$. Hence it can be transformed to a normal form $\Sigma^H_{\bar{2}}$ where the input/output map is linear by the means of the regular static state feedback ${w}$ in \eqref{lineariningW} and the local diffomorphism $H(\bar{x}): \mathbb{U}_{\bar{x}_e} \rightarrow \mathbb{U}_{\Tilde{x}_e}$ 
\begin{equation}
\begin{split}
\left(\begin{array}{c}
     \zeta_1 \\ \zeta_2 \\
     \eta 
\end{array}\right)&= H(\bar{x}):= \left(\begin{array}{c}
     V_1^0\\ E_1+M_b^{-1}D_1 u \\
     \Phi(\bar{x}) 
\end{array}\right)
\label{control_diff}
\end{split}
\end{equation} 
where $\Phi(\bar{x}): \mathbb{U}_{\bar{x}_e}\rightarrow \mathbb{T}^6 \times \mathbb{R}^{6}$ is a sufficiently smooth function chosen such that, in the open neighbourhood $\mathbb{U}_{\bar{x}_e}$, $H(\bar{x})$ is a diffomorphism and $\frac{\partial\Phi}{\partial \Bar{x}}\Bar{G}_2=0$ holds. A possible choice of $\Phi$ is the states of the passive joints, as suggested in~\cite{Spong}, $\Phi:=(\eta_1,\eta_2)^T=({\color{black}{q_p}},{\color{black}{\dot{q}_p}})^T$.

Hence the internal dynamics can be expressed as 
\begin{equation}
\begin{split}
    \dot{\eta}_1&= {\eta}_2\\
    \dot{\eta}_2&=F_{\eta}({\eta}_1,{\eta}_2,{\zeta}_1,\zeta_2,\Bar{z},C^0_1)\\&:={\color{black}{L_{1}}}-M_{pp}^{-1} M_{bp}^T \zeta_2+ M_{pp}^{-1} D_3 \Bar{z}
\end{split} \label{internal_dynamics}
\end{equation}
where $\eta$-dynamics represent the internal dynamics and $D_3$, $\Bar{z}$ are calculated from
\begin{equation}
\begin{split}
        \Bar{z}&=\left(\begin{array}{ccc}
         z_2 & 
            z_3& 
            z_4 
    \end{array}\right)^T\\
    D_3&=\operatorname{blkdiag}\big((k_{2} s_{22},0)^T,(k_{3} s_{32},0)^T,(k_{4} s_{42},0)^T\big)
\end{split}
\label{thrustZD}
\end{equation}
{\color{black}{where $s_{jj}$ and $c_{jj}$ are $\sin(q_{jj})$ and $\cos(q_{jj})$, respectively.}} Furthermore, we can map the controller states to the new coordinates  $(z,{\color{black}{q_a}})^T=f_c({\zeta},\eta,C^0_1)$ where $f_c$ is a solution of the algebraic equations
\begin{equation}
  Q = \Tilde{D} \Bar{z} + {\color{black}{L_3}} z_5
\end{equation}
where $\Tilde{D}, {\color{black}{Q}}, {\color{black}{L_3}}$ are given by 
\begin{equation}
\begin{split}
   \Tilde{D}&\resizebox{0.95\columnwidth}{!}{$
 = \operatorname{Ad}^T_{C^0_{1}}\left(\begin{array}{ccc}
          N_2\operatorname{Ad}^T_{(C^0_{f_2})^{-1}}\Bar{f}_2 & N_3\operatorname{Ad}^T_{(C^0_{f_3})^{-1}}\Bar{f}_3 &N_4\operatorname{Ad}^T_{(C^0_{f_4})^{-1}}\Bar{f}_4
     \end{array}\right)$} \\
   {\color{black}{Q}}&= \operatorname{Ad}^T_{C^0_{1}} M_b(\zeta_2-E_1)\\
    {\color{black}{L_3}} &=\left(\begin{array}{c} {k}_{5}\,s_{52}-\frac{a\,c_{51}\,c_{52}}{2}\\ \frac{c_{52}\,\left(a\,c_{51}-2\,{k}_{5}\,s_{51}\right)}{2}\\ \frac{a\,\left(s_{52}+c_{52}\,s_{51}\right)}{2}+{k}_{5}\,c_{51}\,c_{52}\\ s_{52}\\ -c_{52}\,s_{51}\\ c_{51}\,c_{52} \end{array}\right) 
\end{split}
\end{equation}
Let $\Bar{z}=\Bar{f}_c({\zeta},\eta,C^0_1)$ where $\Bar{f}_c$ is the first 3 elements in the map ${f}_c$. Hence, substituting back into $\eta$-dynamics in~\eqref{internal_dynamics} yields the normal form $\Sigma^H_{\bar{2}}$
\begin{equation}
\Sigma^H_{\bar{2}}:\left\{\begin{split}
   \dot{\zeta}_1&=\zeta_2\\
    \dot{\zeta}_2&=v \\
    \dot{\eta}_1&= {\eta}_2\\
    \dot{\eta}_2&={\color{black}{L_{1}}}-M_{pp}^{-1} M_{bp}^T \zeta_2+ M_{pp}^{-1} D_3 \Bar{f}_c\\
    h_y&={\zeta}_1
\end{split}\right. \label{normal_form}
\end{equation} This proves Statement 1.
\subsubsection{Statement 2 Proof}
$\Sigma_3$ in the new coordinates after applying $H(\bar{x})$  \eqref{control_diff} becomes this $\Sigma^H_3$
\begin{equation}
\Sigma^H_3:\;\;\dot{C}^0_1=[\zeta_1]{C}^0_1
    \label{Csystem}
\end{equation}
Thus, $\Sigma^H_{\bar{2}}$~\eqref{normal_form} interconnected with $\Sigma^H_3$~\eqref{Csystem} is equivalent to the system~\eqref{statespace} with a linear relationship between virtual input $v$ and output from the base spatial twist $V^0_1$.

 The zero dynamics of the network ($\Sigma^H_{\bar{2}}$, $\Sigma^H_3$) are the internal dynamics when $h_y(t) \equiv 0\, \forall t\geq0$. This constraint is imposed for all times by setting the virtual input $v(t)=0\, \forall t\geq0 $ and the initial conditions of $\zeta$-dynamics to be $(\zeta_1(0),\zeta_2(0))=(0,0)$. As a result, the linear differential equations of $\zeta$ admit the solution $(\zeta_1(t),\zeta_2(t))=(0,0)\, \forall t\geq0$. This also leads to $\Sigma^H_3$~\eqref{Csystem} having a solution ${C}^0_1(t)= {C}^0_1(0) \,\forall t\geq 0$. Thus, the zero dynamics manifold $\mathbb{Z}$ is then obtained as
 \begin{equation}
 \resizebox{1\columnwidth}{!}{$
 \begin{split}
          \mathbb{Z}&=\Bigg\{{\eta}(t) \in \mathbb{T}^6 \times \mathbb{R}^{6},\,t\in \mathbb{R}_{\geq0} 
   \,\Bigr\rvert \dot{\eta}(t)=\left(\begin{array}{c}
         {\eta_2}(t) \\
         F_{\eta}\big({\eta}_1(t),{\eta}_2(t),0,0,{C}^0_1(0),\Bar{f}_c(0,\eta(t),{C}^0_1(0))\big)
   \end{array}\right) \Bigg\}
 \end{split}$}
 \label{Z_normal}
 \end{equation}
\begin{remark}
In the original coordinates of system  $(\Sigma_{\Bar{2}}, \Sigma_3)$, one can express the zero dynamics manifold \eqref{Z_normal} by transforming back the the initial conditions and the input constraint $v=0$ from the normal form into the $\Bar{x}$ coordinates. Hence, $\zeta_1(0)={\color{black}{V_1^0}}(0)=0$ while the constraint $\zeta_2(0)=\dot V_1^0(0)=0$ restricts the initial conditions of the dynamic extension states $z(0)^T$ and ${\color{black}{q_a}}(0)$ to be a solution of~\eqref{accZD} with $\dot{V}^0_1(0)=0$. Let us call this solution $f_{c_x}$, then the zero dynamics manifold in the original coordinates $\mathbb{Z}_x$ is defined by
 \begin{equation}
 \resizebox{1\columnwidth}{!}{$\begin{split}
          \mathbb{Z}_x&=\Bigg\{\bar{x}(t) \in \bar{\mathbb{X}},\,t\in \mathbb{R}_{\geq0} 
   \,\Bigr\rvert \Sigma_{\Bar{2}}(\bar{x};\Tilde{w},{\color{black}{C_1^0}}(0)),  \,\Tilde{w}= -D_2^{-1}M_bE_2,\\ &\quad {\color{black}{C_1^0}}(0)\in SE(3), {\color{black}{V_1^0}}(0)=0, (z(0),{\color{black}{q_a}}(0))^T=f_{c_x}\Bigg\}
 \end{split}$}
  \label{ZeroDOriginal}
 \end{equation}
\end{remark}
What is remaining to complete the proof is to study the stability of the zero dynamics. When we restrict $\eta\in\mathbb{Z}$ we get,
 \begin{equation}
     \resizebox{1\columnwidth}{!}{$
     \begin{split}
         \dot{\eta}_1&= {\eta}_2\\
    \dot{\eta}_2&= \Tilde{M}_{pp}^{-1}\big(-D_f {\eta}_1-\operatorname{col}\big(\Tilde{S}_{j}^T(\Tilde{M}^0_{B_j}\dot{\Tilde{S}}_{j}+\dot{\Tilde{M}}^0_{B_j}\Tilde{S}_{j}){\eta_2}_{j}\big) \\&+{\color{black}{{\Tilde{S}}_p^T}} \operatorname{col}({\Tilde{M}}_{B_j}G^0)+ D_3 \Tilde{z}\big)\\
    &= \Tilde{M}_{pp}({\eta}_1)^{-1}\big(-D_f {\eta}_2 - \Tilde{C}({\eta}_1,{\eta}_2){\eta}_2-{\color{black}{\Tilde{g}_p}}({\eta}_1)+ D_3({\eta}_1) \Tilde{z}\big)
     \end{split}$} \label{ZeroD}
 \end{equation}
 The superscript $\Tilde{(.)}$ donates the value of the corresponding variable evaluated at $({C}^0_1(t),\zeta_1(t),\zeta_2(t))=({C}^0_1(0),0,0)$. $\Tilde{C}$ is the Coriolis matrix which is linear in passive joint velocities ${\eta}_2$ and given by
 \begin{equation}
  \resizebox{1\columnwidth}{!}{$
  \begin{split}
          \Tilde{C}=&\operatorname{blkdiag}\big(\Tilde{C}_2(\eta_{1_{2}},\eta_{2_{2}}),\Tilde{C}_3(\eta_{1_{3}},\eta_{2_{3}}),\Tilde{C}_4(\eta_{1_{4}},\eta_{2_{4}})\big)\\
          \Tilde{C}_j=& \left(\begin{array}{cc}
               S_{j1}^T M^0_{B_j}\dot{S}_{j1}-S_{j1}^T \operatorname{ad}_{S_{j2}}^T M^0_{B_j} S_{j1} \eta_{2_{j2}} & S_{j1}^T M^0_{B_j}\dot{S}_{j2}-S_{j1}^T \operatorname{ad}_{S_{j2}}^T M^0_{B_j} S_{j2} \eta_{2_{j2}} \\
  S_{j2}^T M^0_{B_j}\dot{S}_{j1}-S_{j2}^T \operatorname{ad}_{S_{j1}}^T M^0_{B_j} S_{j1}\eta_{2_{j1}} & S_{j2}^T M^0_{B_j}\dot{S}_{j2}-S_{j2}^T \operatorname{ad}_{S_{j1}}^T M^0_{B_j} S_{j2}\eta_{2_{j1}}
          \end{array}\right)
 \end{split}$} \label{Coro}
 \end{equation}
 where $j=2,3,4$. 
 
 A suitable choice of positive definite smooth Lyapunov candidate is the total energy of the rigid bodies connected by the passive joints comprising the internal dynamics $U_\eta(\eta): \mathbb{Z} \rightarrow \mathbb{R}_{>0}\,\forall \eta \neq \eta_{e},\, {\color{black}{{U_\eta}}}(\eta_{e})=0$ $\forall \eta_{e} \in\mathbb{D}^{[i]}_{x_e}$
 \begin{equation}
      {\color{black}{{U_\eta}}}=\frac{1}{2} \eta_2^T \Tilde{M}_{pp}(\eta_1-\eta_{1e})  \eta_2+\Tilde{U}(\eta_1-\eta_{1e}) 
 \end{equation}
 where $\Tilde{U}(\eta_1-\eta_{1e})$ is the potential energy of the passive dynamics in the manifold $\eta\in\mathbb{Z}$ whose time derivative is
 \begin{align}
\dot{\Tilde{U}}&=\eta_2^T\frac{\partial \Tilde{U} }{\partial (\eta_1-\eta_{1e})}\\
          &=  -\eta_2^T{\Tilde{S}_p^T(\eta_1-\eta_{1e})} \operatorname{col}({\Tilde{M}_{B_j}(\eta_1-\eta_{1e})}G^0)=\eta_2^T {\color{black}{\Tilde{g}_p}} \nonumber
 \end{align}where $G^0=\operatorname{col}(0_5, -9.81\,m/s^2)$. Taking the time derivatives of $ {\color{black}{{U_\eta}}}$ yields
 \begin{equation}
 \begin{split}
      {\color{black}{{\dot U_\eta}}}&=\eta_2^T \Tilde{M}_{pp} \dot{\eta}_2+\frac{1}{2}\eta_2^T \dot{\Tilde{M}}_{pp} {\eta}_2+\dot{\Tilde{U}}\\
     &= \eta_2^T(\frac{1}{2} \dot{\Tilde{M}}_{pp}-\Tilde{C}){\eta}_2 \\
     &- \eta_2^T D_f {\eta}_2-\eta_2^T {\color{black}{\Tilde{g}_p}}+\eta_2^T D_3 \Tilde{z}+\eta_2^T {\color{black}{\Tilde{g}_p}}
 \end{split}
 \end{equation}
 Since $\frac{1}{2} \dot{\Tilde{M}}_{pp}-\Tilde{C}$, with $\dot{\Tilde{M}}_{pp}$ and  $\Tilde{C}$ respectively given in Sec.~\ref{app:ModelDefinitions} and~\eqref{Coro}, is skew-symmetric, ${\color{black}{{\dot U_\eta}}}$ reduces to
 \begin{equation}
 \begin{split}
     {\color{black}{{\dot U_\eta}}}&= - \eta_2^T D_f {\eta}_2+\eta_2^T D_3 \Tilde{z}
 \end{split}
 \end{equation}
Thus for asymptotic stability of $\eta_{e}$, $\Tilde{z}$ must satisfy 
\begin{equation}
    \begin{split}
       \eta_2^T D_3 \Tilde{z}<  \eta_2^T D_f {\eta}_2 
    \end{split}
\end{equation}
 Since the quadratic form $\eta_2^T D_f {\eta}_2$ can be bounded by$$\lambda_{\text{min}}(D_f) \|\eta_2\|^2 \leq \eta_2^T D_f \eta_2 \leq \lambda_{\text{max}}(D_f) \|\eta_2\|^2$$
Applying Cauchy–Schwarz inequality to bound $|\eta_2^T D_3\Tilde{z}|\leq\|\eta_2\|\,\|D_3\Tilde{z}\|$ yields the bounds on $\Tilde{z}$ $\forall {\eta} \in \mathbb{Z}-\{\eta_e\}$ to be \eqref{FinalBoundsZD} in original coordinates, where also this norm property $\|D_3\Tilde{z}\|\leq \|D_3\| \|\Tilde{z}\|$ is used. The proof is concluded.
\subsection{Proof of Proposition 2}
\label{proofProp2}
We construct the state feedback for the virtual input $v$ which locally asymptotically stabilizes $\Sigma_{{\zeta}}:\dot \zeta_1=\zeta_2,\,\,\dot \zeta_2=v$ composed with $\Sigma^H_3$ at $(\zeta_{1e},\zeta_{2e},{\color{black}{C^0_{1e}}})$. Thus, from Theorem 10.3.1 in~~\cite{isidori1999nonlinear}, the cascaded interconnection of the autonomous system $\Sigma_{{\zeta}}$ composed with $\Sigma^H_3$, and the internal dynamics $\Sigma_\eta$ is guaranteed to be LAS in $\mathbb{U}_{\bar{x}_e}\times SE(3)$, given the result on the stability of the zero dynamics stated in Theorem~\ref{Theorm1}.
 
 Consider the system $\Sigma^H_3$\eqref{Csystem}. The goal is to find a feedback law for the input $\zeta_1$ to stabilize the equilibrium $C^0_{1e}$. To this end, we choose the positive definite smooth Lyapunov candidate ${\color{black}{U_b}}(C^0_1): SE(3) \rightarrow \mathbb{R}_{>0}\,\forall C^0_1 \neq C^0_{1e},\, {\color{black}{U_b}}(C^0_{1e})=0$ to be  
    \begin{align}
                {\color{black}{U_b}}(C^0_1(t))&= \frac{1}{2} ||C^{1e}_0C^0_1(t)-I||_F^2\\&
                =\frac{1}{2}\operatorname{tr}\big((C^{1e}_0C^0_1(t)-I)^T(C^{1e}_0C^0_1(t)-I)\big) \nonumber
    \end{align}
    where $||.||_F$ is the Frobenius norm and $\operatorname{tr}(.)$ denotes the trace. The term $C^{1e}_0C^0_1(t)$ is the left-trivialized error on $SE(3)$. Define the variable $E_c$ as 
    \begin{equation}
        E_c=C^{1e}_0C^0_1(t)-I \label{EcEqua}
    \end{equation}
    
    Taking the time derivatives of the ${\color{black}{U_b}}(C^0_1)$ along the trajectories of $C^0_1$ while simplifying using properties of the trace yields
\begin{align}
            {\color{black}{\dot U_b}}(C^0_1(t))&=\frac{1}{2} \frac{d}{dt}\operatorname{tr}\big(E_c^TE_c\big)
            =\frac{1}{2} \operatorname{tr}\big(\frac{d}{dt}E_c^TE_c+E_c^T\frac{d}{dt}E_c\big) \nonumber\\
            &=\frac{1}{2} \operatorname{tr}\big(\frac{d}{dt}E_c^TE_c\big)+\frac{1}{2} \operatorname{tr}\big(E_c^T\frac{d}{dt}E_c\big)\nonumber\\
             &=\frac{1}{2} \operatorname{tr}\big(E_c^T\frac{d}{dt}E_c\big)^T+\frac{1}{2} \operatorname{tr}\big(E_c^T\frac{d}{dt}E_c\big)\nonumber\\
             &= \operatorname{tr}\big(E_c^T\frac{d}{dt}E_c\big)
        \end{align}
    Since $\frac{d}{dt}E_c$ is given by
    \begin{equation}
        \dot{E}_c=C^{1e}_0\frac{d}{dt}C^0_1(t)=C^{1e}_0[\zeta_1]{C}^0_1(t) \label{ECdot}
    \end{equation}
    Hence 
\begin{align}
            {\color{black}{\dot U_b}}(C^0_1(t))&= \operatorname{tr}\big(E_c^TC^{1e}_0[\zeta_1]{C}^0_1(t)\big) =\operatorname{tr}\big({C}^0_1(t) E_c^TC^{1e}_0[\zeta_1]\big)\nonumber \\
            &=\operatorname{tr}\big(((C^{1e}_0)^TE_c ({C}^0_1)^T )^T[\zeta_1]\big)
\end{align}
    By using the fact that if $[b] \in se(3), \, A \in \mathbb{R}^{4\times4}$ then this trace property holds $\operatorname{tr}\big(A^T\,[b]\big)=\operatorname{tr}\big(P^T_{se(3)}(A)\,[b]\big)$ where $P_{se(3)}(A)$ projects the matrix $A$ on $se(3)$ as follows
    \begin{equation}
       \resizebox{0.85\columnwidth}{!}{$ P_{se(3)}(A)=\left(\begin{array}{cc}
             \frac{1}{2}(A_{[1:3,1:3]}-A^T_{[1:3,1:3]})  & A_{[1:3,4]}\\
             0_{1\times3} &0
        \end{array}\right)$} \label{ProjectionMatr}
    \end{equation}
    where $A_{[a:b,c,d]}$ constructs a new matrix by extracting from matrix A the rows number $a$ to $b$ and columns number $c$ to $d$, ${\color{black}{\dot U_b}}$ hence becomes
         \begin{equation}
        \begin{split}
            {\color{black}{\dot U_b}}(C^0_1(t)) 
            &=\operatorname{tr}\big(P^T_{se(3)}((C^{1e}_0)^TE_c ({C}^0_1)^T) [\zeta_1]\big)\\
            &=\hat{P}^T_{se(3)}((C^{1e}_0)^TE_c ({C}^0_1)^T)L\zeta_1 
        \end{split}
    \end{equation}
   where $L=\operatorname{diag}(2,2,2,1,1,1)$. Now let the input $\zeta_1$ be
\begin{equation}
    \zeta_1= - K_p \hat{P}_{se(3)}\Big((C^{1e}_0)^TE_c ({C}^0_1)^T\Big)=: \zeta_{1e}
    \label{zeta_1e}
\end{equation}
 where the diagonal gain matrix $K_p>0 \in \mathbb{R}^{6\times6}$ and $\hat{.}:se(3) \rightarrow \mathbb{R}^6$ denotes the inverse of the isomorphism $[.]$. Substituting $\eqref{zeta_1e}$ makes ${\color{black}{\dot U_b}}$ take this quadratic form
    \begin{equation}
         \resizebox{0.86\columnwidth}{!}{$\begin{split}
            {\color{black}{\dot U_b}}(C^0_1(t)) 
            &=-\hat{P}^T_{se(3)}((C^{1e}_0)^TE_c ({C}^0_1)^T)LK_p \hat{P}_{se(3)}((C^{1e}_0)^TE_c ({C}^0_1)^T)\\
            &=-||\big[K_p^{\frac{1}{2}}\hat{P}_{se(3)}((C^{1e}_0)^TE_c ({C}^0_1)^T)\big]||_F^2 <0 \,\,\,\forall E_c\neq0
        \end{split}$}
    \end{equation}
    
Now, we follow a backstepping strategy to stabilize the $\zeta_1$-dynamics at the equilibrium $\zeta_{1e}$~\eqref{zeta_1e} using the input $\zeta_2$. Consider the positive definite smooth Lyapunov candidate ${\color{black}{U_{\zeta_1}}}({C}^0_1,\zeta_1): SE(3) \times \mathbb{R}^{6} \rightarrow \mathbb{R}_{>0},\,\forall C^0_1 \neq C^0_{1e}$ and $ \,\forall\zeta_1\neq\zeta_{1e}, \, \text{ with }{\color{black}{U_{\zeta_1}}}(C^0_{1e},\zeta_{1e})=0$
\begin{equation}
    {\color{black}{U_{\zeta_1}}}({C}^0_1,\zeta_1)= {\color{black}{U_b}}+\frac{1}{2}||\zeta_1-\zeta_{1e}||^2 
\end{equation}
Taking the time derivatives of ${\color{black}{U_{\zeta_1}}}$ gives 
\begin{equation}
    \begin{split}
        {\color{black}{\dot U_{\zeta_1}}}({C}^0_1,\zeta_1)&= \dot{V}_1+(\dot{\zeta}_1-\dot{\zeta}_{1e})^T (\zeta_1-\zeta_{1e})
    \end{split}
\end{equation}
hence by substituting the dynamics of $\zeta_1$~\eqref{normal_form}, i.e., $\dot{\zeta}_1=\zeta_2$ 
\begin{equation}
    \begin{split}
        {\color{black}{\dot U_{\zeta_1}}}({C}^0_1,\zeta_1)&= \dot{V}_1+(\zeta_2-\dot{\zeta}_{1e})^T (\zeta_1-\zeta_{1e})
    \end{split}
\end{equation}
and selecting the input $\zeta_2$ as
\begin{equation}
    \zeta_2=\dot{\zeta}_{1e}+\Bar{\zeta}_2
\end{equation}
thus the derivatives of ${\color{black}{U_{\zeta_1}}}$ reads as
\begin{equation}
    \begin{split}
        {\color{black}{\dot U_{\zeta_1}}}({C}^0_1,\zeta_1)&= \dot{V}_1+\Bar{\zeta}_2^T (\zeta_1-\zeta_{1e})
    \end{split}
\end{equation}
Setting $\Bar{\zeta}_2$ to be $\Bar{\zeta}_2=-K_d({\zeta}_1-\zeta_{1e})$, where the diagonal gain matrix $K_d>0 \in \mathbb{R}^{6\times6}$, renders ${\color{black}{\dot U_{\zeta_1}}}$ negative definite $\forall$ $(C^0_{1},\zeta_{1})\neq(C^0_{1e},\zeta_{1e})$
\begin{equation}
    \begin{split}
        {\color{black}{\dot U_{\zeta_1}}}({C}^0_1,\zeta_1)&= \dot{V}_1-({\zeta}_1-\zeta_{1e})^TK_d (\zeta_1-\zeta_{1e}) <0 
    \end{split}
\end{equation}
The total control law for ${\zeta}_2$ is
\begin{equation}
\begin{split}
     \zeta_2=&\dot{\zeta}_{1e}-K_d({\zeta}_1-\zeta_{1e})=: \zeta_{2e}\\
\end{split}
    \label{zeta_2e}
\end{equation}
  
This step is repeated once again to design a controller $v$ that stabilizes the $\zeta_2$-dynamics~\eqref{normal_form} at the equilibrium $\zeta_{2e}$ with the Lyapunov candidate ${\color{black}{U_{\zeta_2}}}({C}^0_1,\zeta_1,\zeta_2)$ given by
\begin{equation}
    {\color{black}{U_{\zeta_2}}}({C}^0_1,\zeta_1,\zeta_2)={\color{black}{U_b}}({C}^0_1)+{\color{black}{U_{\zeta_1}}}({C}^0_1,\zeta_1)+\frac{1}{2}||\zeta_2-\zeta_{2e}||^2
\end{equation}
The equation for the virtual input $v$ is thus obtained as
\begin{equation}
\begin{split}
        v=&\dot{\zeta}_{2e}-K_a({\zeta}_2-\zeta_{2e})\\
\end{split}
    \label{virtual_input}
\end{equation}
 where $K_a$ is a diagonal gain matrix $K_a>0 \in \mathbb{R}^{6\times6}$. 
 
The time derivatives of ${\zeta}_{1e}$ are functions of the first time derivative of the error $E_c$~\eqref{EcEqua}, i.e. $\dot{E}_c$~\eqref{ECdot}, and
 \begin{align}
     \ddot{E}_c&= C^{1e}_0\Big([\zeta_2]+[\zeta_1]^2\Big){C}^0_1
 \end{align}
 Hence 
\begin{equation}
    \resizebox{0.95\columnwidth}{!}{$\begin{split}
     \dot{\zeta}_{1e}&= - K_p \hat{P}_{se(3)}\Big((C^{1e}_0)^T\Big(\dot{E}_c({C}^0_1)^T+E_c({C}^0_1)^T[\zeta_1]^T\Big)\Big) \\
     \ddot{\zeta}_{1e}&=- K_p \hat{P}_{se(3)}\Big((C^{1e}_0)^T\Big(\ddot{E}_c({C}^0_1)^T+2\dot{E}_c({C}^0_1)^T[\zeta_1]^T+E_c({C}^0_1)^T(([\zeta_1]^T)^2+[\zeta_2]^T)\Big)\Big)
 \end{split}$}
 \label{timeDerizeta1e}
\end{equation}
After some algebraic manipulations of~\eqref{zeta_2e} and~\eqref{virtual_input}, the virtual control $v$ can be rewritten as
\begin{equation}
v=\ddot{\zeta}_{1e}+\bar{K}_d(\dot{\zeta}_{1e}-{\zeta}_{2})+\bar{K}_a({\zeta}_{1e}-{\zeta}_{1})
    \label{virtual_input_final}
\end{equation}
where $\bar{K}_a>0\in \mathbb{R}^{6\times6},\,\bar{K}_d>0\in \mathbb{R}^{6\times6}$ are diagonal gain matrices. They are related to ${K}_a$ and ${K}_d$ by 
\begin{equation}
        \bar{K}_a={K}_d\,{K}_a,\,\, \bar{K}_d={K}_d+{K}_a
\end{equation}
If the error states are defined as
\begin{equation}
    e_{s1}(t):={\zeta}_{1e}-{\zeta}_{1},\,\, e_{s2}(t):=\dot{e}_{s1} \label{ErrorSysProp}
\end{equation}then a linear error system with a state $e_{s}=(\,\,e_{s1}\,\, e_{s1}\,\,)^T$ and a pair $(A,B)$ given in Proposition~\ref{Prop2} results. The gain matrices $\bar{K}_d$ and $\bar{K}_a$ can thus be tuned to exponentially stabilizes its origin by rendering $(A-BK)$ Hurwitz.
\begin{remark}
    Observe that to eliminate the steady-state error in the convergence of $e_{s}(t)$ to zero, which may be present due to parametric uncertainty, an integral term can be added to $v$ as
\begin{equation} 
\resizebox{0.90\columnwidth}{!}{$\begin{split}
    v=\ddot{\zeta}_{1e}+\bar{K}_d(\dot{\zeta}_{1e}-{\zeta}_{2})+\bar{K}_a({\zeta}_{1e}-{\zeta}_{1})+K_I \int ({\zeta}_{1e}-{\zeta}_{1})\, dt
\end{split}$}
    \label{virtual_input_final_integral}
\end{equation}
where ${K}_I={K}^T_I>0\in \mathbb{R}^{6\times6}$ is the integral gain matrix.
\label{RemarkIntegral}
\end{remark}
 This concludes the proof.


\begin{figure*}
   \centering
\begin{subfigure}{0.475\linewidth}
\includegraphics[width=1\columnwidth]{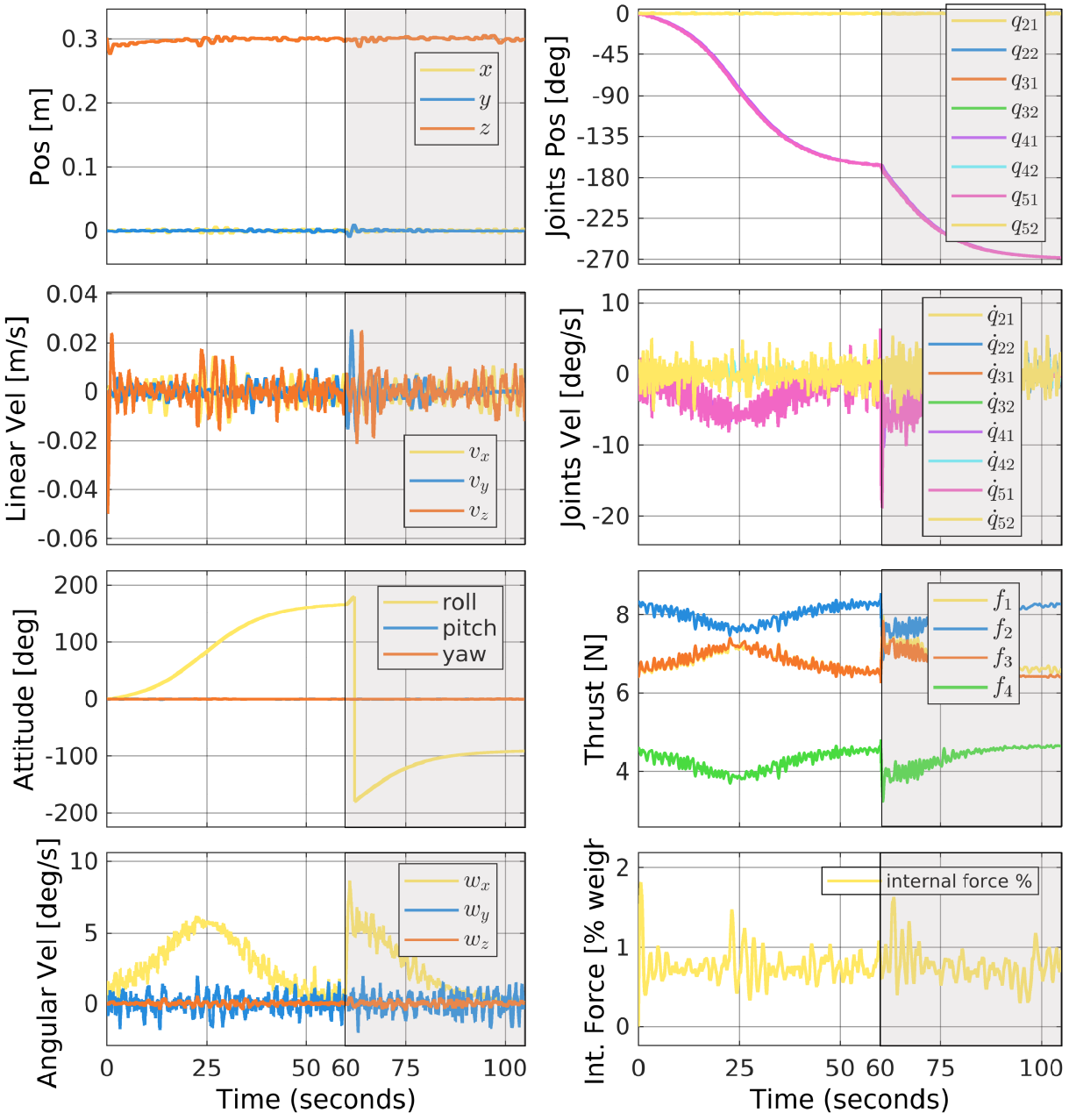}
\caption{PJD$_1$}
\label{Sim_PJD1_270}
\end{subfigure}
\begin{subfigure}{0.475\linewidth}
\includegraphics[width=1\columnwidth]{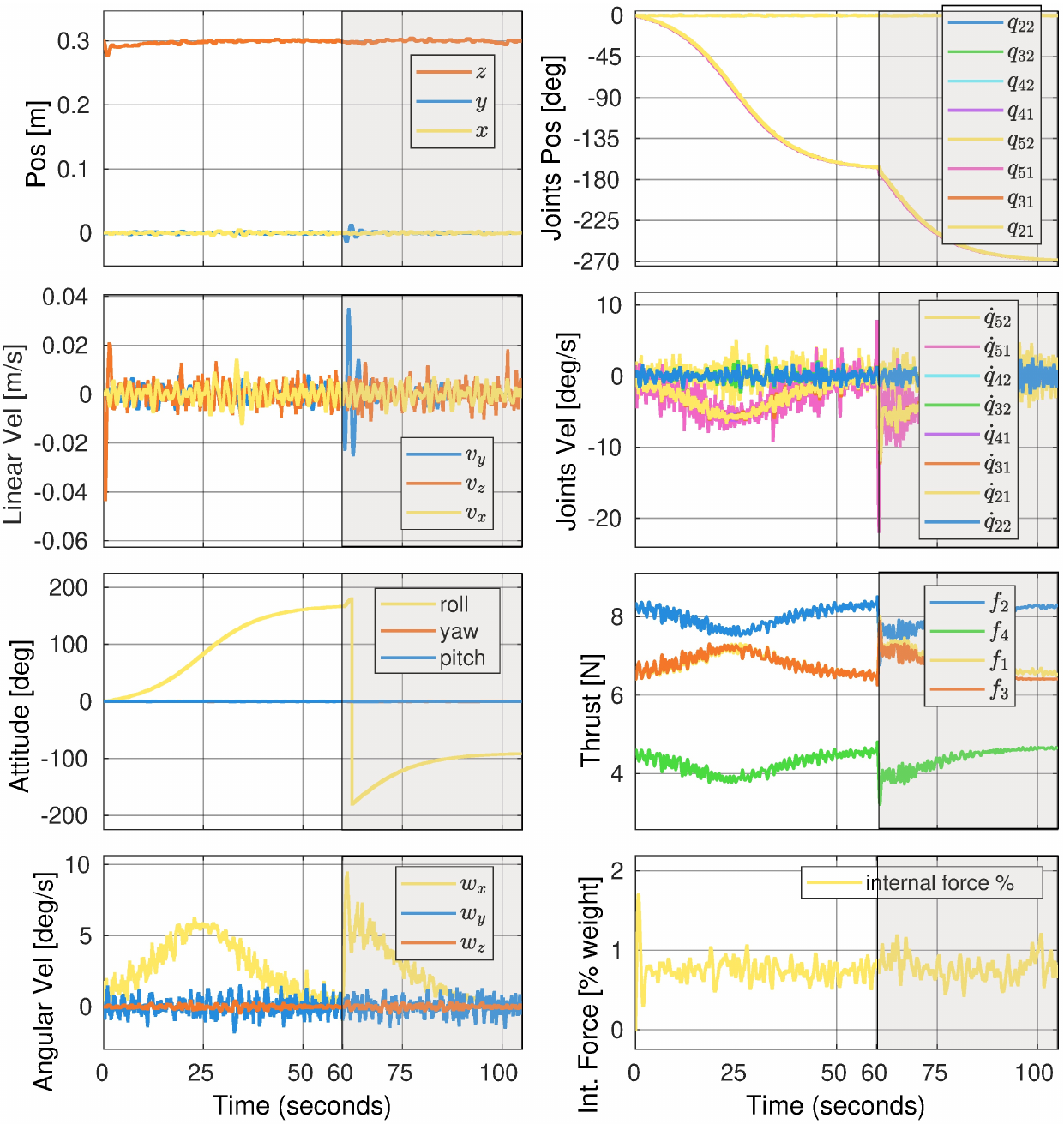}
\caption{PJD$_2$}
\label{Sim_PJD2_270}
\end{subfigure}
\setlength{\belowcaptionskip}{-8pt}
    \caption{{\color{black}{Evolutions of the state and input for (a) PJD$_1$ and (b) PJD$_2$ designs in 270-degree roll maneuver shown in Fig. \ref{fig:3DComplete}. The simulation has two phases: First (0-60s) is the $170^\circ$ roll step with the initial position held constant. Second (60-105s, gray-shaded), the roll attitude step is increased to $270^\circ$. 
    }}}
    \label{Sim270}
\end{figure*}
\end{document}